\documentclass[pre,12pt, a4paper, onecolumn, secnumarabic, amssymb, amsfonts, amsmath, bibnotes, superscriptaddress, floatfix, altaffilletter, eqsecnum, final, nofootinbib]{revtex4}
\usepackage{fullpage}
\usepackage{bm} % Boldmath - allows bold maths symbols
\usepackage{color}
\usepackage{hyperref}
\usepackage{appendix}
\usepackage{graphicx}
\usepackage{enumerate}
\usepackage[english, british]{babel}
\usepackage{float}
\usepackage{eso-pic}
\usepackage{epsfig}
\usepackage[all]{hypcap}
\usepackage{paralist}
\usepackage{subfig}
\usepackage{datetime}
\usepackage{alltt}
\begin{document}
\title{\Large{\center{The Holographic Dual of $2+1$ Dimensional QFTs with $\mathcal{N}=1$ SUSY and Massive Fundamental Flavours}}\\}
\author{Niall T. Macpherson\footnote{\href{mailto:pymacpherson@swansea.ac.uk}{\tt pymacpherson@swansea.ac.uk}}}
\affiliation{Department of Physics, Swansea University\\Singleton Park, Swansea SA2 8PP, United Kingdom}
%% Abstract
%%
\begin{abstract}
\vspace{3.0 cm}
%\begin{center}
%    \normalsize\textbf{}
%\end{center}
\vspace{0.4 cm}
\normalsize{\center{$\bm{Abstract:}$}\vspace{3 mm}\\
The Maldacena Nastase solution is generalised to include massive fundamental matter through the addition of a flavour profile. This gives a holographic dual to $\mathcal{N}=1$ SYM-CS with massive fundamental matter with a singularity free IR. We study this solution in some detail confirming confinement and asymptotic freedom. A recently proposed solution generating technique is then applied which results in a new type-IIA supergravity solution. In a certain limit, the geometry of this solution is asymptotically $AdS_4\times Y$, where Y is the metric at the base of the Bryant-Salamon $G_2$ cone, which has topology $S^3\times S^3$.}\\
\vspace{2.0 cm}
\newpage
\end{abstract}
\vspace{5.0 cm}
\maketitle

% Define the Table of contents
\def\tocname{\Large{Table of contents}}

 %blank page
%\newpage
%\thispagestyle{empty}
%\mbox{}

\setlength{\parskip}{1.5mm}
\newpage
\tableofcontents
\cleardoublepage

% blank page
%\newpage
%\thispagestyle{empty}
%\mbox{}

%\twocolumngrid
%\newpage
%\pagestyle{myheadings}
%\markright{}
%\setlength{\parskip}{0mm}

%% *** *** *** *** *** *** *** *** *** *** *** ***
%%
%% Main body
%%
%% *** *** *** *** *** *** *** *** *** *** *** ***

\newpage

\section{Introduction}
Almost 40 years ago $^\prime$t Hooft proposed the large $N_c$ expansion for strongly coupled gauge theories in the seminal paper \cite{CERN-TH-1786}. This suggested that the large $N_c$ limit of gauge theories would be a type of string theory, however it was not for many more years that this observation would start to bare significant fruit with the advent of the AdS-CFT correspondence \cite{hep-th/9711200}. The correspondence is a weak coupling to strong coupling duality which, in its best understood form, relates $\mathcal{N}=4$ SYM to Type IIB string theory on an $AdS_5\times S^5$ geometry (see \cite{hep-th/9905111} for a review). Developments in integrability on both sides of the correspondence \cite{Serban:2010sr} have shown that it stands up to many no trivial tests and by now it is a well trusted, if not formally proven, tool used to probe the strong coupling dynamics of both gauge and string theories. 

Despite its robust nature the original AdS-CFT correspondence gives a gravity dual of a gauge theory with maximal SUSY and conformal symmetry with only adjoint fields. $\mathcal{N}=4$ Super Yang-Mills is not much like Yang Mills so for phenomenological reasons it was desirable to extend the idea to include non-conformal theories with minimal SUSY or even no SUSY at all. This form of gauge gravity correspondence certainly stands on less stable ground but the belief is that, none the less, it gives useful insights into strongly coupled dynamics. The canonical example of such a gravity dual is the Maldacena-Nunez solution \cite{hep-th/0008001} which is conjectured to be dual in the infra red to pure $\mathcal{N}=1$ SYM in $3+1$ dimensions. A great deal of work has been done on this background so as to give a dual that describes minimal SQCD in $3+1$ dimensions as closely as possible. A major step was to add fundamental matter, the method by which this is achieved consists of adding flavour branes to the gravity dual, which is equivalent to including an open string sector, and was proposed in \cite{hep-th/0205236}. Massless fundamental flavours were added to the Maldacena-Nunez solution, first in the quenched approximation where the number of flavour branes is small enough to neglect their back-reaction \cite{hep-th/0311201}, and later in the in the unquenched Veneziano limit, where $N_c$ and $N_f$ are taken to be infinite but their ratio is finite \cite{hep-th/0602027}\cite{HoyosBadajoz:2008fw}. This was implemented by means of a smearing procedure, proposed in \cite{hep-th/0409133}\cite{hep-th/0505140}, in which the flavour branes are continuously distributed across their common co-dimensions. Unfortunately most gravity duals with massless unquenched fundamental matter are afflicted by non physical IR singularities. The problem is that massless flavours correspond to branes that reach the origin and so an infinite number of back-reacting branes can give rise to infinite curvature in the IR. This was resolved for $3+1$-dimensional SQCD in \cite{Conde:2011rg}, by giving the branes a density profile which increases from $0$ to $1$ as one flows from the IR to the UV, a more phenomenologically motivated approach was proposed in \cite{Barranco:2011vt} but both are equivalent to adding massive fundamental matter. 

In this paper the goal is to continue a parallel story, that of gravity duals which flow in the IR to $\mathcal{N}=1$ SYM in 3-dimensions where it is also possible to have a Chern-Simons term of level $k$. The original dual was proposed in \cite{Schvellinger:2001ib} but was found independently by Maldacena and Nastase in \cite{Maldacena:2001pb} soon after, where the interpretation is clearer. It consists of D5 branes which are extended along the space time directions and wrap a non compact 3-cycle inside a Manifold of $G_2$-holonomy, the solution is reviewed in \autoref{section:MN}. This was conjectured to be dual to the 3-dimensional $\mathcal{N}=1$ Yang-Mills Chern-Simons theory which was studied by Witten in \cite{Witten:1999ds}. Witten calculated the index of this theory and showed that for $k=\frac{N_c}{2}$ the theory exhibits a single confining vacuum. 

Massless fundamental flavours were added to the Maldacena-Nastase solution by Canoura, Merlatti and Ramallo in \cite{Canoura:2008at}. They found two distinct classes of UV consistent with the 3-dimensional IR, those with an asymptotically linear dilaton and those with a asymptotically constant dilaton where the metric approaches a deformed $G_2$-cone. This solution is reviewed in appendix \ref{massless} while the procedure for adding unquenched flavours is reviewed in \autoref{section:AddingFlavour}. The massless flavour background suffers from the IR pathology common to these backgrounds and in fact an IR expansion has not been found. In this paper massive fundamental flavours are added to the Maldacena-Nastase background and we find a solution that interpolates between the deformed  unflavoured background and the massless flavoured backgrounds of \cite{Canoura:2008at}. This is achieved by introducing a profile function $P$ that depends on the holographic coordinate. To determine a suitable $P$ we use the more phenomenologically motivated methods of \cite{Barranco:2011vt} because lacking the partial integration of the BPS equations that so drastically simplifies the 4 dimensional case, it is not clear how to proceed in the more formal fashion of \cite{Conde:2011rg}.

Recently a solution generated method called "rotation" has been introduced \cite{Maldacena:2009mw}. This is a powerful technique that, starting from a relatively simple type-IIB supergravity solution with characterised by an $SU(3)$ structure, enables a more complicated IIB solution to be generated, often with different field content. This technique can be applied to the deformed Maldacena-Nunez solution of \cite{hep-th/0602027} and generates the Baryonic branch \cite{Butti:2004pk} deformation of Klebanov-Strassler \cite{Klebanov:2000hb}. In \cite{Gaillard:2010gy} Gaillard and Martelli introduced a new avatar of rotation which can be implemented on type-IIA supergravity solutions. After an S-duality, this is applied to the massive flavour deformation of the Maldacena-Nastase solution to generate the $G_2$ analogue of the Baryonic branch of Klebanov-Strassler, but with a profile.
  
The lay out of the paper is as follows: An ansatz for the massive flavour gravity dual is laid out in \autoref{section:Ansatz} which is shown in \autoref{section:Solutions} to lead to a non-singular IR. The asymptotic series solutions for this system are consistent deformations of both the unflavoured and massless flavour cases in the appropriate limits which shows that an interpolation between both is possible. In \autoref{section:TestProfile} a profile is proposed on phenomenological grounds to achieve this interpolation and we explore some of the consequence of the addition of massive flavour on the dual field theory in \autoref{section:QFT}. The rotation is described then implemented in \autoref{section:IIA}, where we also attempt to get a handle on the likely form of the field theory dual. Finally the work is summarised and comments made in \autoref{section:diss}

\section{The dual of pure $\mathcal{N}=1$ SYM in 2+1 dimensions}\label{section:MN}
The holographic dual of $2+1$ dimensional $\mathcal{N}=1$ super Yang Mills was found by Maldacena and Nastase in \cite{Maldacena:2001pb} and is based on the 5d supergravity solution of \cite{Chamseddine:2001hk} which can be lifted first to 7d then to 10d. The Maldacena-Nastase background is generated by $N_c$ $D5$ branes that wrap a non compact 3-cycle inside the internal space, which is a manifold of $G_2$ holonomy. In the IR the the supergravity theory reduces to the $2+1$ dimensional theory considered by Witten in \cite{Witten:1999ds}. This theory includes a Chern-Simons coupling $k$ and has a unique ground state if $k = \frac{N_c}{2}$. In this section we review this holographic dual, the notation we use is based on that of \cite{Canoura:2008at}. The 10-d Metric in Einstein frame is expressed, setting $\alpha'g_s = 1$, as:
\begin{equation}\label{eq:MNmetric}
ds^2=  e^{\phi/2} \left(dx^2_{1,2}+dr^2 +\frac{e^{2h}}{4}(\sigma^i)^2 +\frac{N_c}{4} (\omega^i-A^i)^2\right) 
\end{equation}
Where $\sigma^i$ and $\omega^i$ ($i=1,2,3$) are SU(2) left invariant 1-forms which obey the following differential relations:
\begin{equation}
d\sigma^i = -\frac{1}{2}\epsilon_{ijk}\sigma^j\wedge\sigma^k;~~~~~~~~~~d\omega^i = -\frac{1}{2}\epsilon_{ijk}\omega^j\wedge\omega^k
\end{equation}
And the 1-form gauge field $A^i$ is given by:
\begin{equation}\label{eq:MNforms}
A^i = B^i = \frac{1+w(r)}{2}\sigma^i
\end{equation}
This type IIB supergravity theory has a non trivial RR 3-form:
\begin{equation}\label{eq:MNF3}
F_3 = N_c\left(- \frac{1}{4}\bigwedge_{i=1}^3 (\omega^i-B^i) +\frac{1}{4}F^i\wedge(\omega^i-B^i) + H\right)
\end{equation}
Where $F^i$ is the 2-form field strength of $B^i$:
\begin{equation}\label{eq:FieldStrength}
F^i =dB^i +\frac{1}{2}\epsilon_{ijk}B^j\wedge B^k
\end{equation}
And $H$ is a 3-form chosen such that $dF_3=0$. It is easy to check that $H$ satisfies:
\begin{equation}\label{eq:diffH}
dH=\frac{1}{4}F^i\wedge F^i
\end{equation} 
And so H must take the value:
\begin{equation}
H = \frac{1}{32}\frac{1}{3!} \left(2  w^3-6 w + 8\kappa\right)\epsilon_{ijk}\sigma^i\wedge\sigma^j\wedge\sigma^k
\end{equation}
From the prospective of \autoref{eq:diffH}, $\kappa$ is an integration constant, however it is in fact related to the Chern-Simmons coupling, $k= N_c\kappa$. The value of $\kappa$ can be determined by the fact that the pull back of $F_{(3)}$ onto the cycle on which the color branes are wrapped must vanish in the IR for the background to be non singular. In what follows it is useful to introduce, as in \cite{Canoura:2008at}, the following holographic coordinate:
\begin{equation}
\rho=e^{2h}
\end{equation}
The cycle must shrink to zero as $\rho\to0$, which defines the IR, so that we are left with fractional D2-branes there. As pointed out in \cite{Canoura:2008at} the appropriate cycle is:
\begin{equation}\label{eq:cy}
\Sigma=\left\{\sigma^i|\omega^i=\sigma^i\right\}
\end{equation}
Which has the following induced metric:
\begin{equation}
ds_{\Sigma}^2= \frac{e^{\phi/2}}{4}\left[\rho +\frac{(1-w)^2}{4}N_c\right](\sigma^i)^2
\end{equation}
Clearly $ds_{\Sigma}^2\to0$ for $\rho \to 0$ as long as the dilaton is finite at the origin as $w_{\text{IR}}=1$. The Pull back of $F_3$ on $\Sigma$ is given by:
\begin{equation}
\frac{N_c}{4}\left(\kappa-\frac{1}{2}\right)\sigma^1\wedge\sigma^2\wedge\sigma^3
\end{equation}
Thus clearly $F_3$ vanishes on $\Sigma$ for all values of the holographic coordinate provided
\begin{equation}
\kappa=\frac{1}{2}
\end{equation}
In the IR the two 3-spheres should be manifestly disentangled so that it is possible to factorise the directions parallel and orthogonal to the brane volume in a well defined way. The mixing of the spheres is controlled by the gauge field $A^i$, such that if it becomes zero the spheres are manifestly separated. It is easy to show that when $w=1$ the field strength of $A^i$ is vanishing which implies that $A^i$ is pure gauge and can thus be set to zero via a suitable gauge transformation. This confirms the choice:
\begin{equation}
w_{\text{IR}}=1
\end{equation}
The solution to this theory is given, in the IR, by the following expansion of background fields $w$ and $\phi$: 
%\begin{equation}\label{eq:MNSolution}
%\begin{split}
%\rho = &e^{2h}\\
%w(\rho) = &1 -\frac{1}{3 N_c}\rho + \frac{1}{36 N_c^2}\rho^2 +\frac{1}{216 N_c^3}\rho^3 + ...\\
%\phi(\rho) = &\phi_0 + \frac{7}{24N_c}\rho + \frac{41}{1728 N_c^2}\rho^2 + ...
%\end{split}
%\end{equation}
\begin{equation}\label{eq:MNSolution}
\left.
  \begin{array}{l l}
  \vspace{3 mm}
w(\rho) = &1 -\frac{1}{3 N_c}\rho + \frac{1}{36 N_c^2}\rho^2 +\frac{1}{216 N_c^3}\rho^3 + ...\\  
\vspace{3 mm}
\phi(\rho) = &\phi_0 + \frac{7}{24N_c}\rho + \frac{41}{1728 N_c^2}\rho^2 + ...\\
  \end{array}\right.
\end{equation}
In \cite{Canoura:2008at} it was shown that there is in fact a family of backgrounds related to the Maldacena-Nastase solution. If we generalise the metric \autoref{eq:MNmetric} with a deformation such that $N_c \to F(\rho)$ and instead of the choices of 1-form in \autoref{eq:MNforms} we take:
\begin{equation}\label{eq:1forms}
A^i = \frac{1+w(r)}{2}\sigma^i~~~~~~~~~~B^i = \frac{1+\gamma(r)}{2}\omega^i
\end{equation}
from which it follows that $H$ is now given by
\begin{equation}
H = \frac{1}{32}\frac{1}{3!} \left(2  \gamma^3-6 \gamma + 8\kappa\right)\epsilon_{ijk}\sigma^i\wedge\sigma^j\wedge\sigma^k
\end{equation}
So that the cycle defined by \autoref{eq:cy} is still vanishing in the IR it is clear that $F_{\text{IR}}$ must be a finite constant which shall be referred to as $F_0$.  With the IR conditions we have thus far derived it is simple to show (the details are in \cite{Canoura:2008at}) that $\gamma$ must also satisfy the following for a finite dilaton:
\begin{equation}
\gamma_{\text{IR}}= 1
\end{equation}
Then this set-up solves the BPS equations and supergravity equations of motion with the following IR expansions of the background fields in terms of $\rho=e^{2h}$:
\begin{equation}\label{eq:MNSolution2}
\left.
  \begin{array}{l l}
  \vspace{3 mm}
   			  F(\rho)=  &F_0+\frac{  \left(F_0-N_c\right) \left(9 F_0+5
   N_c\right)}{12 F_0^2}\rho + \frac{ \left(F_0-N_c\right) \left(19 F_0 N_c^2-4 F_0^2 N_c+36 F_0^3+23N_c^3\right)}{144 F_0^5}\rho ^2 + ...\\
   
\vspace{3 mm}

     		     \gamma(\rho)=  &1 - \frac{1}{3 F_0}\rho + \frac{4 N_c F_0-4 N_c^2 + F_0^2}{36 F_0^4} \rho^2+...\\ 
     			
\vspace{3 mm}
     			
    	w(\rho)=	&1 +\frac{\left(2 N_c-3 F_0\right)}{3F_0^2}\rho + \frac{18N_c^3-19 F_0 N_c^2-16F_0^2 N_c+18 F_0^3}{36 F_0^5}\rho^2 + ...\\

  \vspace{3 mm}

       \phi(\rho)=  &\phi_0 + \frac{7 N_c^2 }{24 F_0^3}\rho + 
\frac{N_c^2 (335 N_c^2 - 168 N_c F_0 - 126 F_0^2)}{1728 F_0^6}\rho^2+ ...\\
  \end{array} \right.
\end{equation}
%\begin{equation}\label{eq:MNSolution2}
%\begin{split}
%\rho = &~~e^{2h}\\
%F(\rho) = & F_0+\frac{  \left(F_0-N_c\right) \left(9 F_0+5
 %  N_c\right)}{12 F_0^2}\rho +\\
  % 		 &\frac{ \left(F_0-N_c\right) \left(19 F_0 N_c^2-4 F_0^2 N_c+36 F_0^3+23
   %N_c^3\right)}{144 F_0^5}\rho ^2 + ...\\
%w(\rho) = & 1 +\frac{
 %  \left(2 N_c-3 F_0\right)}{3
  % F_0^2}\rho + \\
   %&\frac{
   %N_c^3-19 F_0 N_c^2-16
   %F_0^2 N_c+18 F_0^3}{36 F_0^5}\rho^2 + ...\\
%\gamma(\rho) = &1 - \frac{1}{3 F_0}\rho + \frac{4 N_c F_0-4 N_c^2 + F_0^2}{36 F_0^4} \rho^2+...\\
%\phi(\rho) = & \phi_0 + \frac{7 N_c^2 }{24 F_0^3}\rho + 
% \frac{N_c^2 (335 N_c^2 - 168 N_c F_0 - 126 F_0^2)}{1728 F_0^6}\rho^2+ ...
%\end{split}
%\end{equation}
The Maldacena-Nastase solution \autoref{eq:MNSolution} is just the special case given by $F_0 =N_c$ were $\gamma=w$. The IR of this family of backgrounds, which will henceforth be referred to as deformed Maldacena-Nastase, is non-singular with any finite choice of $F_0$.

\section{Addition of Unquenched Flavour}\label{section:AddingFlavour}
From the perspective of a gravity dual, $N_f$ unquenched flavours can be added to a gauge theory by the addition $N_f$ flavour branes. These flavour branes are fundamentally different from their color brane counterparts. The $N_c$ coincident color branes give rise to an $SU(N_c)$ gauge symmetry and are converted into a flux. While the $N_f$ coincident flavour branes produce an $SU(N_f)$ global symmetry, these branes will back-react on the geometry of the unflavoured system and modify the metric, BPS equations and equations of motion. Massless flavours are given by branes that reach the origin of the holographic coordinate, while massive flavours by branes which do not.

Usually the $N_f$ flavour branes will be stacked on top of each other with Dirac-delta functions representing their positions in their co-dimensions, this makes the process of finding a back-reacted geometry computationally difficult. A method of getting round this is to use a smearing procedure \cite{hep-th/0409133}\cite{hep-th/0505140}\cite{hep-th/0602027} that de-localises the branes and gives solutions that depend on a single holographic coordinate (See \cite{Nunez:2010sf} for a review). The smearing procedure means that the branes are no longer coincident in their co-dimensions and so generically the global symmetry is broken $SU(N_f)\to U(1)^{N_f}$.

A major advantage of the smearing procedure is that it lends itself well to the application of the powerful techniques of calibrated geometry \cite{hep-th/9905156} (See \cite{hep-th/0305074} for a review and \cite{Gaillard:2008wt} for several examples in the context of smeared flavours)\footnote{These techniques are applicable to non-smeared branes, but lead to partial differential BPS equations}.

To add flavours to the Maldacena-Nastase background (See \autoref{section:MN}) we need to add D5-branes that fill the $2 + 1$ Minkowski dimensions and wrap a non compact 3-cycle in the internal space extending along the radial direction. These flavour branes are smeared with the smearing parametrised by a 4-form $\Omega_s$. The action of this theory will be composed of 2 parts, that of type IIB supergravity and that of flavour branes respectively:
\begin{equation}
S = S_{\text{IIB}} + S_{\text{flavour}}
\end{equation}  
$S_{\text{flavor}}$ includes contributions from both a DBI and a WZ term. Provided the flavour brane, with embedding $X^{a}(\xi)$, satisfies the super symmetry condition:
\begin{equation}\label{eq:SuperCal}
X^*\mathcal{K} = \sqrt{-\hat{g}_6}d^6\xi
\end{equation}
Where $\hat{g}_6$ is the induced metric on the brane, the flavour action can written as a 10-dimensional integral in terms of the smearing form $\Omega_s$ and calibration 6-form $\mathcal{K}$ of the flavour brane as:
\begin{equation}\label{eq:Sf}
S_{\text{flavour}}=-T_5\int_{\mathcal{M}_{10}} \left(e^{\phi/2}\mathcal{K}-C_{(6)}\right)\wedge \Omega_s
\end{equation}
The calibration 6-form can be expressed in terms of the G-2 structure form, $\Phi$ as
\begin{equation}\label{eq:Kal}
\mathcal{K} = e^0\wedge e^1\wedge e^2 \wedge \Phi
\end{equation}
$\Phi$ and the vielbein basis are defined in \autoref{eq:PhiJ} and \autoref{eq:vielbeins} respectively. $\mathcal{K}$ has the same form for both massive and massless flavours (and in their absence) but it will give rise to different BPS equations because of the difference in definitions of the RR 3-form and metric.

The type IIB action, $S_{\text{IIB}}$, contains a term of the form:
\begin{equation}\label{eq:S2B}
-\frac{T_5}{(2\pi)^2}\frac{1}{2} \int_{\mathcal{M}_{10}} e^{-\phi}F_{(7)}\wedge* F_{(7)}
\end{equation}
From \autoref{eq:Sf} and \autoref{eq:S2B} the equation of motion for $C_{(6)}$ can be obtained. It gives rise to a Maxwell equation for $F_{(7)}$:
\begin{equation}
d\left(e^{-\phi}* F_{(7)}\right) =-(2\pi)^2 \Omega_s
\end{equation}
At this point we recall that in the Einstein frame the 3 and 7-form field strengths are related by $* F_{(7)}=-e^{\phi}F_{(3)}$. Thus the addition of massive flavours leads to the following violation of Bianchi identity of the RR 3-form:
\begin{equation}\label{eq:dF3}
d F_{(3)}= 4\pi^2\Omega_s
\end{equation} 
We use \autoref{eq:dF3} as our definition of the smearing form. So when we wish to add flavour to a background we must modify the RR 3-form to account for this.
\section{2+1 Dimensional $\mathcal{N}=1$ SYM with Massive Flavours} 
In this section we add massive fundamental flavours to the Maldacena-Nastase solution.

\subsection{Massive flavour deformation ansatz}\label{section:Ansatz}
In this section we will propose an ansatz for the massive flavour gravity dual of 3-dimensional, $\mathcal{N}=1$ gauge theories. We use the same metric ansatz as massless flavour case, namely:
\begin{equation}\label{eq:Metric}
ds^2 = e^{2f}\left(dx_{1,2}^2+ dr^2+\frac{e^{2h}}{4} (\sigma^i)^2+\frac{e^{2g}}{4}(\omega^i-A^i)^2\right)
\end{equation}
but choose a more general ansatz for the RR 3-form: 
\begin{equation}\label{eq:F3masive}
F_3 = -\frac{N_c}{4} \bigwedge_{i=1}^3 (\omega^i-B^i) +\frac{N_c}{4}\sum_{i=1}^3(F^i+F_f^i)\wedge(\omega^i-B^i) +N_c(H+H_f)
\end{equation}
The 1-forms $A^i$ and $B^i$ are still defined by \autoref{eq:1forms}
and, using this, the 2-form field strength of $B^i$ can be written explicitly as:
\begin{equation}
F^i = \frac{\gamma'}{2}dr\wedge\sigma^i+\frac{\gamma^2-1}{8}\epsilon_{ijk}\sigma^j\wedge\sigma^k
\end{equation} 
The forms $F_f$ and $H_f$ are our flavour deformation forms parametrizing the violation of the Bianchi identity. We chose, as in \cite{Conde:2011rg}, an ansatz which parallels $F_i$ and $H$:
\begin{equation}
\begin{split}
F_f^i = & \frac{1}{2}L_1(r)dr\wedge\sigma^i +\frac{1}{8}L_2(r)\epsilon_{ijk}\sigma^j\wedge\sigma^k\\
H_f = & \frac{1}{32}\frac{1}{3!}L_3(r)\epsilon_{ijk}\sigma^i\wedge\sigma^j\wedge\sigma^k
\end{split}
\end{equation}
The components of $F_3$ which differ from the massless case, \autoref{eq:coF3massless}, are the following:
\begin{equation}\label{eq:coF3}
\begin{split}
F^{(3)}_{ri\hat{i}} &=~\frac{1}{2}N_c 
   \left(L_1+\gamma'\right)e^{-3f-g-h}\\
F^{(3)}_{\hat{i}jk} &= -\frac{N_c}{2}\epsilon_{ijk}\left(1+w^2-L_2-2w\gamma\right)e^{-3f-3h}
\end{split}
\end{equation}
Where V is modified to:
\begin{equation}\label{eq:V}
V = \left(1-w^2\right) (w-3\gamma) -4 \left(1-\frac{3 L_2}{4}\right) w
  +8 \kappa +L_3-3 L_2 \gamma
\end{equation}
From the last line of \autoref{eq:coF3} we can see that for $L_2 = 4\frac{N_f}{N_c}$ and $L_1=0$ we almost reproduce the massless flavoured theory of \cite{Canoura:2008at}, but \autoref{eq:V} spoils this. As we seek a dual to a theory where flavours become massless in the UV this clearly must be reconciled. The first step is to choose:
\begin{equation}
L_3 = 3(\gamma+C )L_2
\end{equation}
Where $C$ is a constant. This allows \autoref{eq:V} to reduce to \autoref{eq:V0} for $L_2=4\frac{N_f}{N_c}$ but also gives the correct term in the flavourless limit $L_2 = 0$ provided $\kappa=\frac{1}{2}$ as in \autoref{section:MN}. This is consistent as we require the theory flows to the deformed Maldacena-Nastase background in the IR. The UV expansions of \cite{Canoura:2008at} will only be reproduced when $C=-1$ (See Appendix \ref{massless}, however the BPS system in Appendix \ref{BPS} is consistent for any value of $C$ and ,for reasons that will become apparent later in this section, it shall be kept arbitrary for now.

The BPS equations impose a restriction on $L_1(r)$, \autoref{eq:dL2Rule}, which implies that $L_1(r)\propto L_2'(r)$. So there is only one function, $L_2(r)$, that parametrises the flavour deformation. Let us define a profile function, $P(r)$, that we want to interpolate between $0$ and $1$ as we run from the IR to the UV:
\begin{equation}\label{equation:P}
P(r) =\frac{N_c L_2(r)}{4N_f}
\end{equation}
The ansatz now depends only on the original background fields and $P$., But we still have an arbitrary constant $C$ that we need to fix. This can be archived if we consider the pull back of the RR 3-form, defined in \autoref{eq:F3masive}, onto various 3-cycles in the geometry.

$F_{(3)}$ obeys the flux quantisation condition\footnote{One must choose a representation for the left invariant one forms ie: $\sigma^1=\cos\psi d\theta +\sin\psi\sin\theta d\phi$, $\sigma^2=-\sin\psi d\theta +\cos\psi\sin\theta d\phi$, $\sigma^3= d\psi+\cos\theta d\phi$ where the angles are defined between $0\leq\theta\leq\pi$, $0\leq\phi<2\pi$ and $0\leq\psi<4\pi$}:
\begin{equation}\label{equation:D5Quantisation}
-\frac{1}{2\kappa^2_{10}T_5}\int_{\tilde{S}^3}F_3 =N_c
\end{equation}
Where $\tilde{S}^3$ is the 3-sphere parametrised by $\omega^i$ and we set $2\kappa^2_{10}=(2\pi)^7$ and $(2\pi)^5T_5=1$. This is of course also true for both the massless flavour and flavourless theories as can be readily seen if we consider the pull-back of $F_3$ onto $\tilde{S}^3$:
\begin{equation}
-\frac{N_c}{4}\omega^1\wedge\omega^2\wedge\omega^3
\end{equation}  
Which, being independent of the profile function must hold for all values of $P$, in particular $P=0$ and $P=1$.

The pull back of $F_{(3)}$ onto the shrinking cycle $\Sigma$ (\autoref{eq:cy}), on which the color branes are wrapped, will still vanish in the IR. However in general, unlike the deformed Maldacena-Nastase solution, it will no longer vanish over the whole range of the holographic coordinate $\rho=e^{2h}$. The pull back onto $\Sigma$ is given by:
\begin{equation}\label{eq:pullF��}
F_{(3)}\bigg\lvert_{\Sigma}=\frac{N_c}{4}\left[\kappa-\frac{1}{2} +\frac{3N_f}{2N_c}(C+1)P\right]\sigma^1\wedge \sigma^2\wedge\sigma^3
\end{equation} 
This confirms that, as with deformed Maldacena-Nastase, we must have that $\kappa=\frac{1}{2}$. We could choose $C=-1$ at this stage to impose that \autoref{eq:pullF��} vanishes for all $\rho$ and, as previously stated, we would reproduce the UV expansions of \cite{Canoura:2008at} when $P=1$. While this may seem attractive, it is not required for the background to be IR finite as $P(\rho=0)=0$. 

To field theory dual will have a Chern-Simons level which can be calculated, as in \cite{Maldacena:2001pb}, by integrating $F_{(3)}$ over a 3-cycle which is non vanishing in the IR. Two such cycles are easy to find, they are the 3-spheres $S^3$ and $\tilde{S}^3$, parametrised by $\sigma^i$ and $\omega^i$ respectively. They have the following induced metrics:
\begin{equation}
ds^2_{S^3}=\frac{1}{4}e^{\phi/2}\left[\rho+\frac{F}{4}(1+w)^2\right](\sigma^i)^2;~~~~~~ds^2_{\tilde{S}^3}=\frac{1}{4}e^{\phi/2}F(\omega^i)^2
\end{equation}
Which are both clearly non zero provided $F(\rho=0)\neq 0$ and $w(\rho=0)\neq -1$, which are already required if we wish to flow to the deformed Maldacena-Nastase theory in the IR. The 3-Sphere $S^3$ is the cycle we need to calculate the Chern-Simmons level, the 3-cycle $\tilde{S}^3$ gives the flux quantisation condition \autoref{equation:D5Quantisation}. If we now consider a probe D5-brane with the embedding $\Xi=(x,y,t,S^3)$, we know that there is a coupling of the form:
\begin{equation}\label{eq:chern}
 -\frac{1}{16\pi^3}\int_{\Xi}F_3\wedge tr[\mathcal{A}\wedge d\mathcal{A} +\frac{2}{3}\mathcal{A}\wedge \mathcal{A}\wedge \mathcal{A}]=  -\frac{\tilde{k}}{4\pi}\int_{\mathbb{R}^{1,2}}tr[\mathcal{A}\wedge d\mathcal{A} +\frac{2}{3}\mathcal{A}\wedge \mathcal{A}\wedge \mathcal{A}] 
\end{equation}
where here, $\mathcal{A}$ refers to a world volume field on the brane and after the equality we have integrated $F_{(3)}$ over $S^3$. On $\Xi$, $F_3$ takes a comparatively simple form because there is no longer any terms proportional to $w^i$ or $dr$:
\begin{equation}\label{eq:cher}
\frac{N_c}{4}\left(1 +\frac{3N_f}{2N_c}(C-1)P\right)\sigma^1\wedge\sigma^2\wedge\sigma^3
\end{equation}
All the dependence of \autoref{eq:chern} on $\sigma^i$ is contained in $F_3$. Therefore it is possible to perform the integral over $S^3$ and be left with a Chern-Simmons term of level $\tilde{k}$ on the remaining dimension of the brane $(t,x,y)$. We see however that \autoref{eq:cher} can only give a quantised $\tilde{k}$ when:
\begin{equation}
C=1
\end{equation}
Although this choice of $C$ will give different UV expansions than those of \cite{Canoura:2008at} (When $P=1$) we will see that they are still perfectly consistent as a supergravity solutions. Indeed when $P=1$ it is argued in appendix \ref{massless} that there is a whole family of consistent UV expansions of which the choice in \cite{Canoura:2008at} is only a particular case. Specifically the choices $\kappa=\frac{1}{2}$ and $C=1$ made here will lead to another specific case of the expansions \autoref{eq:lineardilaton} and \autoref{eq:constantdilaton}, and this choice is the only sensible one when massive flavour is added and we demand that the Chern-Simons level is everywhere quantised. 

Performing the integral over $S^3$ then gives the Chern-Simons level\footnote{This is in stark contrast to the massless flavour case of \cite{Canoura:2008at} where $\tilde{\kappa}=N_c-3N_f$, which may be calculated by setting $P=1$, $C=-1$ in \autoref{eq:cher}. But as $N_f$ always appears with $P$ in the massive case this cannot be quantised}:
\begin{equation}
\tilde{k}=\frac{1}{4\pi^2}\int_{S^3}F_3=N_c
\end{equation}
This is not the whole story. As pointed out in \cite{Maldacena:2001pb}, in the flavourless ($P=0$) case, there are 6 dimensional Kaluza-Klein modes giving rise to massive adjoint gluinos. In the IR these will be integrated out and induce a shift of $-\frac{N_c}{2}$ in $\tilde{k}$. As we have no fundamental matter in the IR, this shift will be reproduced in the massive flavour case. So that in the 3 dimensional Chern-Simons term will be given by:
\begin{equation}
k=\tilde{k}-\frac{N_c}{2}=\frac{N_c}{2}
\end{equation}

It is worth pointing out that we could relax the requirement that $\tilde{k}$ is everywhere quantised. If we don't fix C we are led to $\tilde{k} = N_c +\frac{3N_f}{2}(C-1)P$. This is of course quantised in both the IR and the UV, with $\tilde{k}_{IR} = N_c$ and $\tilde{k}_{UV} = N_c+ \frac{3N_f}{2}(C-1)$. It is only in the intermediate region that $\tilde{k}$ becomes dependent on $r$ and there is a field theory interpretation for this. As we flow from the IR the dependence on $r$ causes an increase in $P$ which can be interpreted as increasing the energy of the dual field theory such that the effects of quarks running in loops becomes increasingly noticeable. When $P=1$ the energy is high enough that the full effect of flavours running in loops is visible, the flavours are effectively massless and $\tilde{k}$ is once more quantised but now dependent on $N_f$. The difference in the $\tilde{k}_{UV}$ and $\tilde{k}_{IR}$ is then a shift induced by integrating out the massive flavours. This is an attractive picture, indeed one expects the Chern-Simons level to depend on $N_f$ when the quarks run in loops and the shift in $\tilde{k}$ could be a sign of the parity anomaly in $2+1$ dimensions. One reason we have chosen to instead impose that $\tilde{k}$ is quantised over the whole range of r is that it is the only way to fix $C$. Another reason is that in \autoref{section:IIA} a solution generating technique is applied to the solution we consider here. This technique seems to generically generate solutions in which the sources no longer act as flavours\cite{Gaillard:2010qg}\cite{Conde:2011aa}. Thus the picture of a change in level due to integrating out massive flavours is invalid in that case. At any rate it seems that the qualitative results of the following sections are insensitive to the choice of $C$. Indeed, there exists asymptotic solutions to the differential BPS equations, \autoref{eq:BPS}, for arbitrary $C$ that are IR finite. The numerical analysis performed in \autoref{section:Matching} and \autoref{section:QFT} has been repeated with $C=-1$ (which leads to the UV of \cite{Canoura:2008at}) and $C=0$ with very similar results.

With the constants determined as $C=1$ and $\kappa=\frac{1}{2}$ we can now explicitly write $L_1$ in terms of $P$. We use \autoref{eq:dL2Rule} to write:
\begin{equation}
L_1(r) = \frac{4N_f}{N_c}\eta(r)P'(r)
\end{equation}
Where for $\kappa=\frac{1}{2}$, $C=1$, $\eta$ reduces to a comparatively simple form:
\begin{equation}\label{eq:eta}
\eta\left.\begin{array}{l l}
= \frac{ N_c \left(e^{2 g} \left(3 (w+1)^2 \gamma-2 w^3-3 w^2 -7\right)+e^{2 h} (4+12 \gamma -8 w)\right)+3 e^{2 g} (w-1) \left(e^{2 g} (w+1)^2+4 e^{2 h}\right)}{3
    \left(e^{2 g} (w+1)^2+4 e^{2 h}\right) \left(e^{2 g} (w-1)^2+4 e^{2 h}-4 P N_f\right)-N_c \left(8 \left(w^2-1\right) e^{2h}+e^{2 g} (w+3) (w-1)^3+16 e^{4 h-2g}\right)}
   \end{array}\right.
\end{equation}
It only remains now to calculate the smearing form, $\Omega_s$. Taking the exterior derivative of \autoref{eq:F3masive} we can express this in terms of $P$ as:
\begin{equation}\label{eq:SmearingForm}
\begin{split}
\Omega_s =
-&\frac{N_f}{8\pi^2}\bigg[\frac{1}{8}P \epsilon_{ilm}\epsilon_{ijk}\sigma^l\wedge\sigma^m\wedge\omega^j\wedge\omega^k+\frac{1}{2}\eta P'\epsilon_{ijk}dr\wedge\sigma^i\wedge\omega^j\wedge\omega^k   \\
&~~~~~~~~~~~~~~~~~~~~~~~~~~~~~~~~~~-\frac{1}{4}(2\eta+1)P'\epsilon_{ijk}dr\wedge\sigma^i\wedge\sigma^j\wedge\omega^k\bigg]
\end{split}
\end{equation}
The first term is present in \autoref{eq:masslessOmega} but the other 2 are new\footnote{In general there is a 4th $C$ dependent term $\frac{N_f(C-1)}{64} \eta P'\epsilon_{ijk}dr\wedge\sigma^i\wedge\sigma^j\wedge\sigma^k$, but for $C=1$ this is zero. The other terms only depend on $C$ through $\eta$ and potentially $P$.} It is now clear that both the unflavoured and massless flavour background are special cases of this more general background. In particular \autoref{eq:SmearingForm} reduces to the smearing form of the massless theory, \autoref{eq:masslessOmega}, when $P=1$.

The ansatz laid out in this section defines a set of BPS equations that are derived in Appendix \ref{BPS}. In \cite{arXiv:0706.1244}
 it was proven that for general type II backgrounds with calibrated brane sources, the source-corrected Einstein and dilaton equations
of motion follow automatically from the BPS equations once the likewise source-corrected form equations of motion and Bianchi identities are imposed. The flavourless limit of our background is the deformed Maldacena-Nastase solution for which it was shown in \cite{Canoura:2008at} that the BPS equations imply that the supergravity equations of motion are satisfied. Therefore we know that for our background the supergravity equations of motion are satisfied.

\subsection{Asymptotic Solutions for General Profiles }\label{section:Solutions}
\autoref{eq:BPS} gives a set of BPS equations in terms of an arbitrary profile function $P$. We seek solutions to the BPS equations that in the IR will be given by \autoref{eq:MNSolution2} but will tend to either \autoref{eq:lineardilaton} or \autoref{eq:constantdilaton} in the UV. The profile function $P$ should interpolate between 0 and 1 as we run from the IR to the UV. Formally its precise form will depend on the specifics of the brane embedding as is the case for the dual $\mathcal{N}=1$ SQCD with massive flavours, \cite{Conde:2011rg}. However from a phenomenological point of view we might choose any $P$ that interpolates between 0 and 1 and obeys some consistency requirements.

To begin with let us keep P general and seek asymptotic expansions. The only restrictions we shall impose is that $P~\epsilon~[0,1]$, that P has a minimum in the IR, a maximum in the UV and that it is monotonic. This is the general idea advocated in \cite{Barranco:2011vt}. As proposed in \cite{Canoura:2008at}, we will attempt to find series solutions to the BPS equations in terms of the holographic coordinate $\rho$. Thus if we define:
\begin{equation}\label{eq:Frho}
\rho=e^{2h};~~~~F=e^{2g}
\end{equation}
where we have that:
\begin{equation}
\frac{d \rho}{dr} = 2h'(r)e^{2h(r)}=2h'(r)\rho~~~ \frac{dF}{dr} =2g'(r)e^{2g(r)}= 2g'(r) F 
\end{equation}
we can use the BPS equations \autoref{eq:BPS} to eliminate $g'(r)$ and $h'(r)$. So the following set of differential equations give the BPS equations in terms of the holographic coordinate $\rho$:
\begin{equation}\label{eq:BPSrho}
\left.\begin{array}{l l}
F'(\rho) = \frac{2g'(r) F}{2h'(r)\rho};&w'(\rho) = \frac{w'(r)}{2h'(r)\rho};\\
\gamma'(\rho)=  \frac{\gamma'(r)}{2h'(r)\rho};&\phi'(\rho) =  \frac{\phi'(r)}{2h'(r)\rho};\\
\end{array}\right.
\end{equation}
Where $\phi = 4 f$ and $w'(r)$, $g'(r)$, $h'(r)$, $f'(r)$ are defined by \autoref{eq:BPS} and \autoref{eq:Frho} is used to eliminate the dependence on $g(r)$ and $h(r)$ in favour of $F$ and $\rho$. It is in fact only the first 3 differential equations in \autoref{eq:BPSrho} that are coupled, once they are solve $\phi'(\rho)$ need only be integrated. 

We expect, as in \cite{Canoura:2008at}, that for $N_c\geq2N_f$ the IR of the theory corresponds to $\rho\approx0$. For we expect that $N_c<2N_f$ $\rho$ is no longer such a good holographic coordinate and the theory will be harder to interpret. Sadly, as with the massless case, an analytic solution to \autoref{eq:BPSrho} valid over the whole range of $\rho$ will not be possible to derive and we will have to satisfy ourself with series solutions in the IR and UV that we will then numerically match. Although the series presented in the following sections are for $C=1$ there also exist similar series leading no IR singularity for arbitrary $C$.

\subsubsection{IR series solution}\label{section:IR}
Now that the IR boundary conditions have been determined and we have fixed the integration constant, $\kappa$ it is possible to solve \autoref{eq:BPSrho} as a series expansion about $\rho=0$. At this stage we will keep the discussion general and propose the following arbitrary IR form for the profile\footnote{We might wish to consider $P=P_{\frac{3}{2}}\rho^{\frac{3}{2}}+P_2\rho^2+P_{\frac{5}{2}}\rho^{\frac{5}{2}}+...$. This also leads to sensible results, but with more complicated solutions to the BPS equations so these extra terms have been omitted here} :
\begin{equation}\label{eq:PIR}
P(\rho) = P_2 \rho^2 + P_3\rho^3 +P_4\rho^4 + ...
\end{equation}  
Where there is no constant or linear term so that the profile vanishes and has a minimum at the origin.
It is possible to show (with Mathematica) that the solution to the first three equations in \autoref{eq:BPSrho} close to zero is given by:
\begin{equation}\label{eq:IRSOL}
\left.
  \begin{array}{l l}
  \vspace{3 mm}
&F(\rho)=F_0 + \frac{(F_0-N_c)(5N_c +9 F_0)}{12F_0}\rho+\bigg[\frac{\left(F_0-N_c\right) \left(-4 F_0^2 N_c+19 F_0 N_c^2+23 N_c^3+36 F_0^3\right)}{144 F_0^5}\\

\vspace{3 mm}
&~~~~~~~~~~~~~~~~-\frac{P_2 N_f \left(F_0 N_c+6 N_c^2-3 F_0^2\right)}{6 F_0^2}\bigg]\rho^2+...\\

\vspace{3 mm}

     		     &\gamma(\rho)=1-\frac{1}{3F_0}\rho+\bigg[\frac{4 F_0 N_c-4 N_c^2+F_0^2}{36 F_0^4}+\frac{(3F_0-4N_c)N_fP_2}{3N_cF_0}\bigg]\rho^2+...\\

\vspace{3 mm}
     			
    &w(\rho) =1+\frac{2N_c-3F_0}{3F_0^2}\rho+\bigg[\frac{18 N_c^3-16 F_0^2 N_c-19 F_0 N_c^2+18 F_0^3}{36 F_0^5}+\frac{N_c N_f P_2}{3F_0^2}\bigg]\rho^2+...\\
  \end{array} \right.
\end{equation}
%\begin{equation}\label{eq:IRSOL}
%\begin{split}
%F(\rho)= & F_0 + \frac{(F_0-N_c)(5N_c +9 F_0)}{12F_0}\rho - \frac{4N_c N_f P_2}{15F_0}\rho^{3/2}+... \\
%          &\big(\frac{N_f P_2}{4}-\frac{(F_0-N_c)(23N_c^3 +19N_c^2 F_0-4N_c F_0^2 +36F_0)}{144F_0^5} +\\
%          & \frac{N_c^2N_f(5-5N_c^2-4N_c F_0+9F_0)P_2}{12F_0} -\frac{2N_f^2 P_2}{25}\big{)}\rho^2 +...\\				
%w(\rho)= & 1 + \frac{2N_c-3F_0}{3F_0^2}\rho +\\
%&\frac{19N_c^2F_0 +16N_cF_0^2 -18(N_c^3+F_0^2)-6N_c(N_c^2-1)N_f F_0^3 P_2 }{36F_0}\rho^2+... \\
%\gamma(\rho) = & 1- \frac{\rho}{3F_0}-\frac{4N_f P_2\rho^{3/2}}{5 N_c}  +...\\
%& \frac{\left(N_c(F_c^2-4N_c F_0 - 4N_c^2)-6N_f F_0^3(F_0+N_c(2-2N_c^2+3N_c F_0)P_2)\right)\rho^2}{36N_c F_0^4} +...
%\end{split}
%\end{equation}
The massive flavour IR expansions now depends on $N_f$, notice however that the first two terms in each of these expansions are exactly that of \autoref{eq:MNSolution2}, the deformed Maldacena-Nastase solution, this explicitly shows that our theory becomes flavourless in the IR. In fact if we were to set $P_i=0$ for all $i$ we would recover all the terms in \autoref{eq:MNSolution2}. 

\autoref{eq:IRSOL} can now be used to obtain $\phi$ via integration:
\begin{equation}
\phi(\rho) = \phi_0 +\bigg[\frac{7 N_c^2}{24 F_0^3}+\frac{3 P_2 N_f}{2}\bigg]\rho +...
\end{equation}
Which clearly gives a non-singular dilaton in the IR however if we want to be sure that the IR of our theory is pathology free we should calculate the curvature invariants. These have the following leading order terms close to zero:
\begin{equation}
\left.
  \begin{array}{l l}
  
\vspace{3 mm}

 R &={\scriptstyle e^{-\frac{\phi_0}{2}}}\bigg[\frac{\left(7 N_c^2+108 F_0^3 P_2 N_f\right)}{6 F_0^3}+{\scriptstyle P_2 N_f }\left({\scriptstyle378 P_2 N_f}-\frac{47 N_c^2}{F_0^3}\right)\bigg]...\\

\vspace{3 mm}

\left(R_{\alpha\beta}\right)^2&={\scriptstyle e^{-\phi_0}}\bigg[\frac{1883 N_c^4}{216 F_0^6}-\frac{47 P_2 N_c^2 N_f}{F_0^3}+{\scriptstyle 378 P_2^2 N_f^2}\bigg]+... \\

\vspace{3 mm}

\left(R_{\alpha\beta\gamma\delta}\right)^2&={\scriptstyle e^{-\phi_0}}\bigg[\frac{1231 N_c^4-2952 F_0^2 N_c^2+3240 F_0^4}{216 F_0^6}+\frac{P_2 \left(15552 F_0^5 N_f-10584 F_0^3 N_c^2 N_f\right)}{216 F_0^6}+{\scriptstyle270 P_2^2 N_f^2}\bigg]+...\\

  \end{array} \right. 
\end{equation}
So the solution to the BPS equations gives a background that has no curvature singularity in the IR and reduces the the Maldacena-Nastase solutions of \autoref{section:MN}. 

\subsubsection{UV series solutions}\label{section:UV}
We now need to find series solutions about infinity that are compatible with those of appendix \ref{massless}, the massless flavour case. We propose the UV profile:
\begin{equation}\label{eq:PUV}
P(\rho)= 1 + \frac{P_1}{\rho}+\frac{P_2}{\rho^2}+...
\end{equation}
There are two cases: The asymptotically linear dilaton and the flavoured $G_2$ cone where the dilaton is asymptotically constant. We can take some guidance from \autoref{eq:constantdilaton} and \autoref{eq:lineardilaton} in finding these solutions as they must be reproduced when all $P_i=0$.

It is possible to show that the asymptotically linear dilaton background is given by the following expansions in the background fields:
\begin{equation}
\left.
  \begin{array}{l l}
  \vspace{3 mm}
F(\rho)= & {\scriptstyle N_c}+\frac{N_c N_f}{\rho}-{\scriptstyle N_cN_f}\frac{N_c-12N_f-4P_1}{4\rho}+\\
		
\vspace{3 mm}

&{\scriptstyle N_cN_f}\frac{21 N_c^2+16 \left(8 P_1 N_f+15 N_f^2+P_2\right)-4 N_c \left(37 N_f+7 P_1\right)}{16\rho^3}+...\\
   
\vspace{3 mm}

\gamma(\rho)=&\frac{N_c+3N_f}{2\rho}+\frac{5(N_c-2N_f)(N_c+3N_f)+12N_fP_1}{8\rho^2}+\\

\vspace{3 mm}

&\frac{\left(N_c+3 N_f\right) \left(49 N_c^2-208 N_c N_f+252 N_f^2\right)+4 \left(11 N_c-72 N_f\right)P_1 N_f +48 N_f P_2 }{32\rho^3}+...\\
  		          		         			
\vspace{3 mm}
     			
w(\rho)=&\frac{N_c+3N_f}{2\rho}+\frac{5(N_c-2N_f)(N_c+3N_f)+12N_fP_1}{8\rho^2}+\\
&\frac{\left(N_c+3 N_f\right) \left(49 N_c^2-184 N_c N_f+204 N_f^2\right)+20 P_1 N_f \left(3 N_c-16 N_f\right)+48 P_2 N_f}{32\rho^3}+ ...\\
  \end{array} \right.
\end{equation}
%\begin{equation}
%\begin{split}
%F(\rho)= & N_c +N_c N_f\frac{1}{\rho} - N_cN_f\frac{3N_c-12N_f-4P_1}{4\rho}+\\
%&N_cN_f\frac{21N_c^2-148N_c N_f+240N_f^2+(128N_f-28N_c)P_1 +16P_2}{16\rho^3}+...\\
%w(\rho)=& \frac{N_c-3N_f}{2\rho}+\frac{5N_c^2-25N_c N_f+6N_f(5N_f-2P_1)}{8\rho^2}+\\
%&\frac{49N_c^3-331N_c^2N_f+4N_cN_f(189N_f-37P_1)+4N_c^{1/2}N_fP_1 -12N_f(51N_f-28N_fP_1+4P_2)}{32\rho^3}+...\\
%\gamma(\rho)=&\frac{N_c-3N_f}{2\rho}+\frac{5N_c^2-25N_cN-f+6N_f(5N_f-2P_1)}{8\rho^2}+\\
%&\frac{49N_c^3-355N_c^2N_f+4N_cN_f(219-37P_1)+12N_c^{1/2}N_fP_1-12N_f(63N_f^2-28N_fP_1+4P_2)}{32\rho^3}+...
%\end{split}
%\end{equation}
The series expansion is of the same general form as \autoref{eq:lineardilaton} with the leading order terms matching exactly, the higher order terms reduce to those of \autoref{eq:lineardilaton} when $P=1$. Notice that there unlike the IR there is no free UV constant. We can then integrate to find the form of the dilaton, it is given by:
\begin{equation}
\phi(\rho)=\frac{\rho}{2N_c-4N_f} -\frac{3N_c^3-12N_cN_f+8N_f(2N_f-P_1)}{8(N_c-2N_f)^2}\text{Log}\rho+O\left(\frac{1}{\rho}\right)
\end{equation}
Exactly in line with our expectations. At this stage we notice that we get completely different UV behaviour depending on whether $N_c\geq2N_f$ or $N_c<2N_f$. It was argued in \cite{Canoura:2008at}, the massless flavour background, that for $N_c<2N_f$ the beta function of the dual field theory develops a Landau pole in the UV while for $N_c\geq2N_f$ the dual theory is asymptotically free - we shall explore these issues in the context of massive flavours in \autoref{section:QFT}.  

It is also possible to find the equivalent series for the asymptotic flavoured $G_2$ cone. They are given by the following series expansion in the background fields:
\begin{equation}\label{eq:G2series1}
\left.
  \begin{array}{l l}
  \vspace{3 mm}
F(\rho) = & \frac{4}{3}\rho + {\scriptstyle4(N_f-N_c)} +\frac{11N_c^2-63N_cN_f-4N_f(3N_f+P_1)}{\rho}+...\\
   
\vspace{3 mm}

\gamma(\rho)= & \frac{1}{3}+\frac{N_f}{N_c} +\frac{N_fP_1}{N_c\rho}+ ...\\  		          		         			
\vspace{3 mm}
     			
w(\rho) = & \frac{3(N_c + 3N_f)}{2\rho} + ...\\  \end{array} \right.
\end{equation}
%\begin{equation}
%\begin{split}
%F(\rho) = & \frac{4}{3}\rho + 4(N_f-N_c) +\frac{11N_c^2-15N_cN_f-4N_f(3N_f+P_1)}{\rho}+...\\
%\gamma(\rho)= & \frac{1}{3}-\frac{N_f}{N_c} -\frac{N_fP_1}{N_c\rho}+\frac{\sqrt{3}N_fP_1}{N_c\rho^{3/2}}+ ...\\
%w(\rho) = & \frac{3(N_c - 3N_f)}{2\rho} + ...\\
%\end{split}
%\end{equation}
These in turn lead to the following form for the dilaton:
\begin{equation}\label{eq:constdil}
\phi(\rho) = \phi_{\infty}-\frac{9N_f}{4\rho}-\frac{9(2N_c+3N_f)+36N_fP_1}{32\rho^2}+...
\end{equation}
Which gives the constant dilaton we expect for background field consistent with the massless asymptotic flavoured $G_2$ cone \autoref{eq:constantdilaton}.

\subsection{A choice of profile}\label{section:TestProfile}
\begin{figure}
 \begin{center}
  \subfloat[~]{\label{fig:arctan}\includegraphics[width=0.3\textwidth]{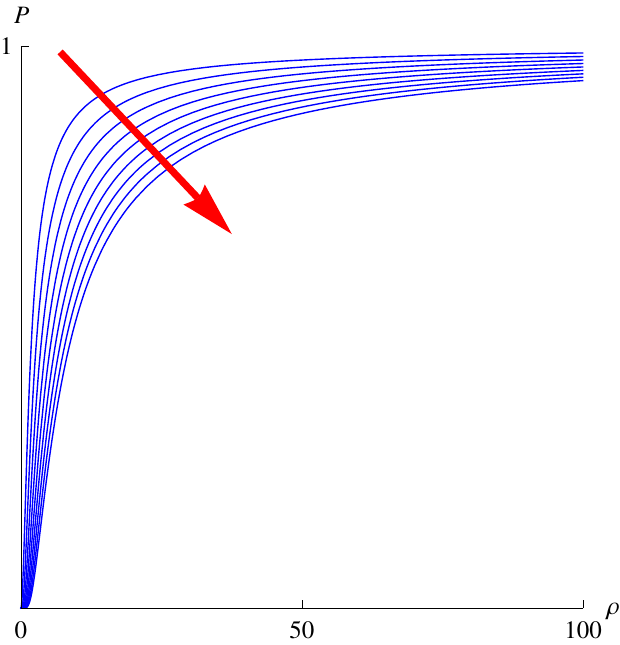}}   ~~~~~~~~~~~~~~~~          \subfloat[~]{\label{fig:darctan}\includegraphics[width=0.3\textwidth]{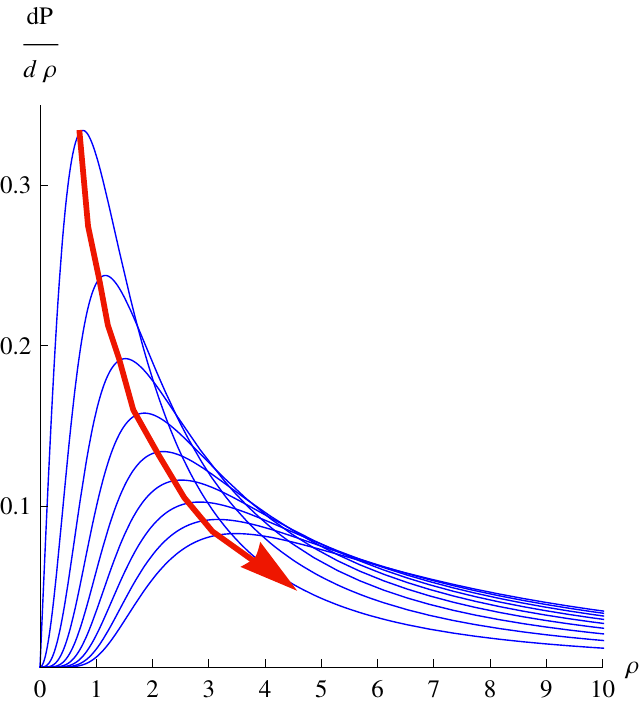}}
  \caption{Graphs of $P(\rho)$ and $P'(\rho)$ for $P(\rho)=\left(\frac{2}{\pi}\arctan\rho\right)^n$. The red arrow indicates the direction of increasing n which takes values between 2 and 10.}
  \label{fig:profile}
 \end{center}
\end{figure}

In the previous section we showed that any choice of $P$ that is monotonically increasing from 0 to 1 as $\rho$ runs from 0 to $\infty$ and which reduces to \autoref{eq:PIR} and \autoref{eq:PUV} in the appropriate limits will give a background that is free from IR pathologies. So it seems that a possible choice for $P$ is the following:
\begin{equation}\label{eq:profile2}
P(\rho)=\left(\frac{2}{\pi}\arctan\rho\right)^n;
\end{equation} 
Where $n\geq2$ is an integer. As can be seen from Fig.\autoref{fig:arctan} this Profile satisfies the requirement of monotonicity with $P\in[0,1]$. It also has the correct form of IR and UV expansion. For $n=6$ we get:
\begin{equation}
\left(\frac{2}{\pi}\arctan\rho\right)^6= \left\{
  \begin{array}{l l}
   \frac{64 \rho ^6}{\pi ^6}-\frac{128 \rho ^8}{\pi ^6}+\frac{2752 \rho ^{10}}{15 \pi ^6}-\frac{43520 \rho ^{12}}{189 \pi ^6}+...& \quad \rho \approx 0\\
  1-\frac{12}{\pi  \rho }+\frac{60}{\pi ^2 \rho ^2} +\frac{4 \left(\pi ^2-40\right)}{\pi ^3 \rho ^3}-\frac{40 \left(\pi ^2-6\right)}{\pi ^4 \rho ^4}+... & \quad \rho \approx \infty\\
  \end{array} \right.
\end{equation}
The choice of $P$ given by \autoref{eq:profile2} will correspond to a brane configuration concentrated around the maximum of $P'$ which we take as a measure of the common quark mass $m_q$. As shown in Fig.\autoref{fig:darctan}, $m_q$ gets larger with $n$ which in turn implies that the range over which there is effectively no flavours increases with $n$ also\footnote{For larger values of $n$ the width of the brane distributions becomes quite large and the idea that most of the quarks in the system have mass $m_q$ becomes invalid}. The UV limit, where the quarks become effectively massless, is defined by $\rho >> m_q$ while the IR limit is given by $\rho<<m_q$, where there are no massive flavours. It turns out that $n=6$ is a good choice for the theory (See \autoref{section:Loops}). In the next section it is confirmed that it is possible to numerically match the IR and UV expansions of \autoref{section:Solutions} with this choice of $P$. 

\subsection{Numerical matching}\label{section:Matching}
\begin{figure}
 \begin{center}
  \subfloat[]{\label{fig:lindil}\includegraphics[width=0.333\textwidth]{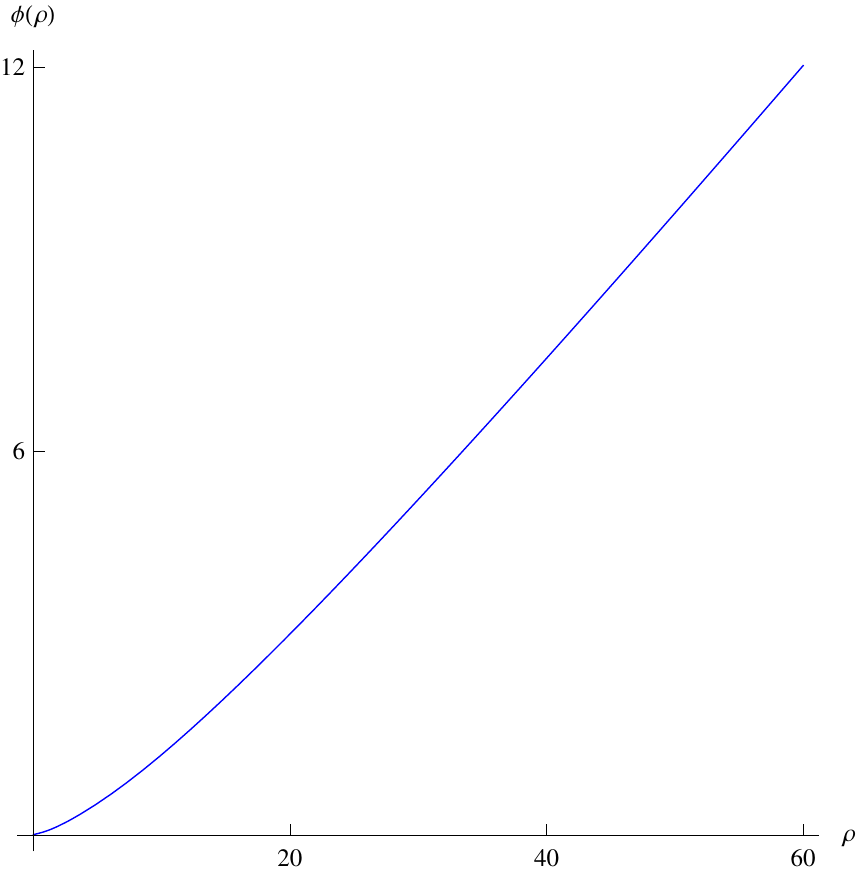}}~~~~~~~~~~~                
  \subfloat[]{\label{fig:constdil}\includegraphics[width=0.333\textwidth]{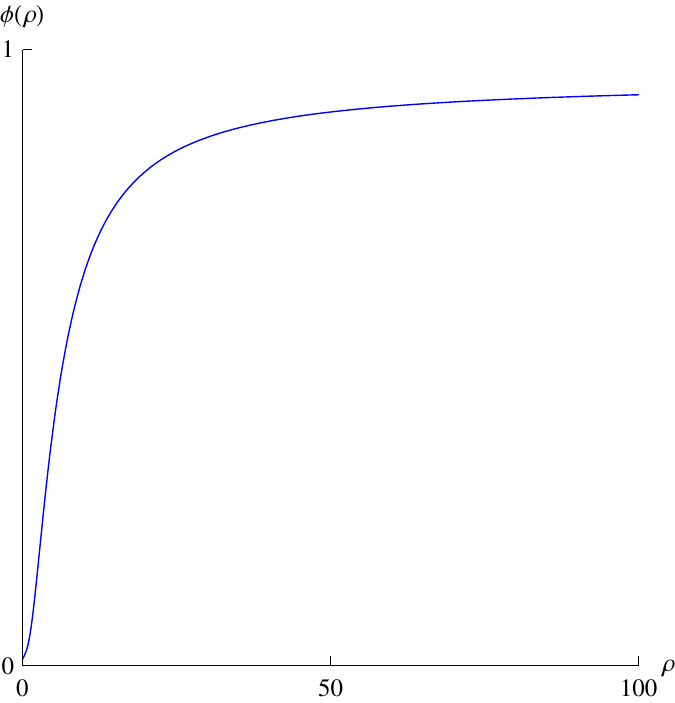}}
  \caption{Numerical graph of the two different dilaton solutions for $N_c=4$ and $N_f=1$. The left graph is asymptotically linear and the right is asymptotically constant.}
  \label{fig:dilatons}
 \end{center}
\end{figure}

In order to numerically integrate the BPS system we must make a concrete choice of profile. Somewhat presciently we choose:
\begin{equation}\label{eq:profile}
P(\rho)=\left(\frac{2}{\pi}\arctan\rho\right)^6;
\end{equation} 
As the numerics work well for this choice. It is now possible to check whether our IR and UV boundary conditions give a solution to the BPS equations that is smooth and continuous over the entire range of the holographic coordinate $\rho$. The standard method employed to achieve this is to use Mathematica (or some other program) to numerically solve the BPS system. This will give interpolating functions for the background fields which will be valid between $\rho \approx 0$ and some finite upper bound $\rho_{max}>>m_q$. The numerical solutions from Mathematica can then be compared to the semi-analytic solutions defined in terms of series about the IR and UV boundaries. If the numerical and semi-analytical results coincides over a significant range of the holographic coordinate we can then say, with confidence, that our background is well defined over the whole space, despite having not rigorously proved this.
\begin{figure}
 \begin{center}
  \subfloat{\label{fig:FUVlindil}\includegraphics[width=0.333\textwidth]{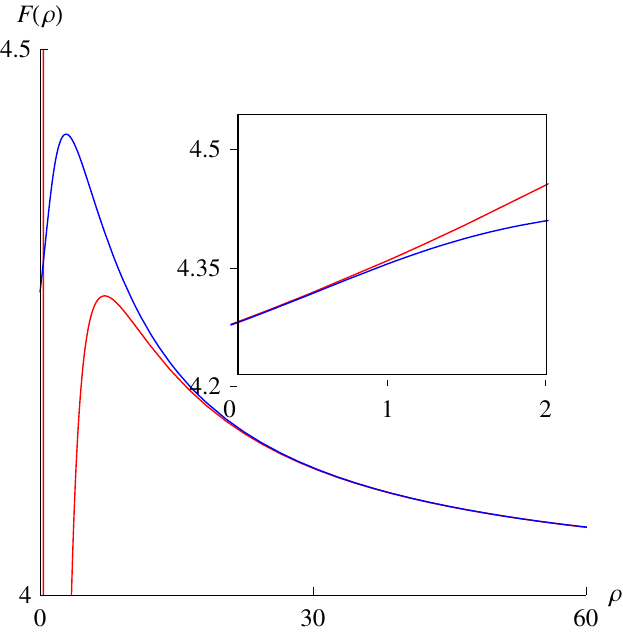}}                
  \subfloat{\label{fig:gUVlindil}\includegraphics[width=0.333\textwidth]{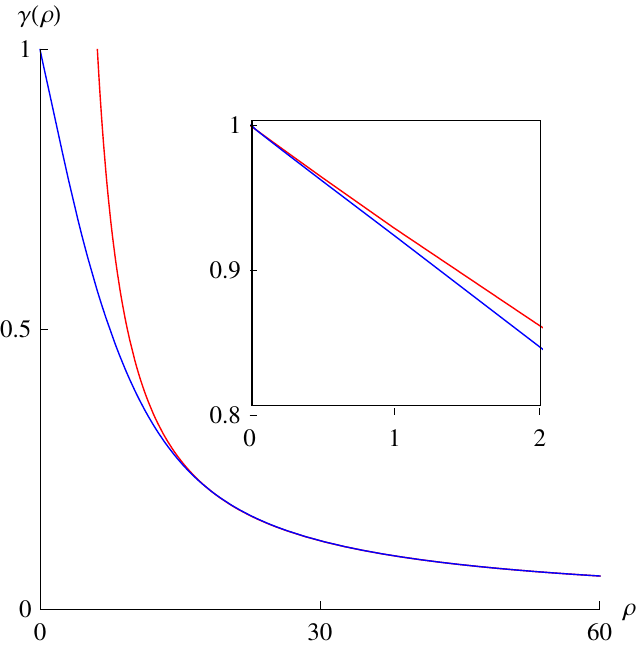}}
  \subfloat{\label{fig:wUVlindil}\includegraphics[width=0.333\textwidth]{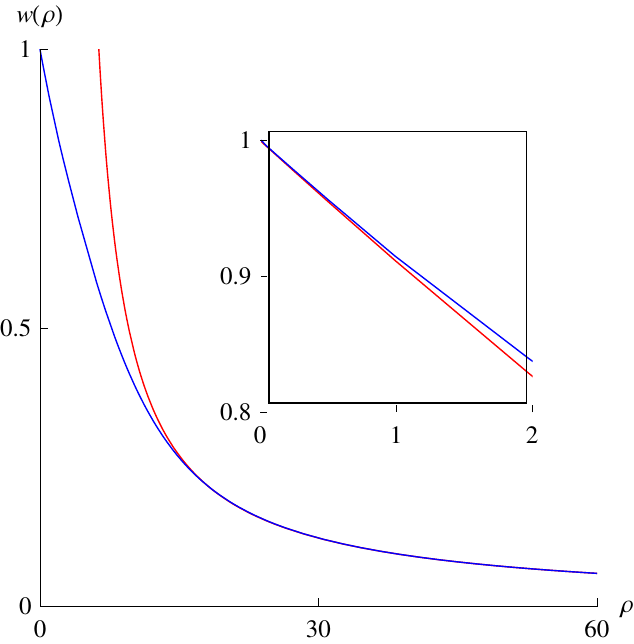}}\\
  \caption{Numerical matching of asymptotically linear dilaton background for $N_c=4$ and $N_f=1$. The red lines
  are graphs of the series solutions of \autoref{section:Solutions}, while the blue lines are numerical solution. The larger graphs show the matching of the UV and the smaller graphs the IR.}
  \label{fig:lindilmatching}
 \end{center}
\end{figure}
Our massive flavour BPS system is defined by \autoref{eq:BPSrho}. By fine tuning the value of $F$ at the origin, $F_0$, it is possible to generate numerical solutions that match both our IR and UV series solutions defined in \autoref{section:IR} and \autoref{section:UV}. \autoref{fig:lindilmatching} shows plots of the background field for the asymptotically linear dilaton background with $N_c=4$ and $N_f=1$. From this we see that the semi-analytic and numerical solution do indeed overlap and as can be seen in Fig.\autoref{fig:lindil} the dilaton does become linear for large $\rho$. In Fig.\autoref{fig:constdil} there is a numerical plot of an asymptotically constant dilaton for $N_c=4$ and $N_f=1$, the corresponding numerical background fields are shown to match their semi-analytic counterparts in \autoref{fig:G2matching}. Although we have not found the explicit form of $F_0$ it does depend on $N_c$, $N_f$ and the IR cut off. $F_0$ needs to be highly fine tuned to match the IR expansions with the asymptotically linear dilaton UV. If $F_0$ is increased from this fine tuned value
there are many numerical solutions that give asymptotically constant dilaton backgrounds and far less tuning is needed to match IR expansions to the UV expansions.
\begin{figure}
 \begin{center}
  \subfloat{\label{fig:FNcNf41G2UV}\includegraphics[width=0.333\textwidth]{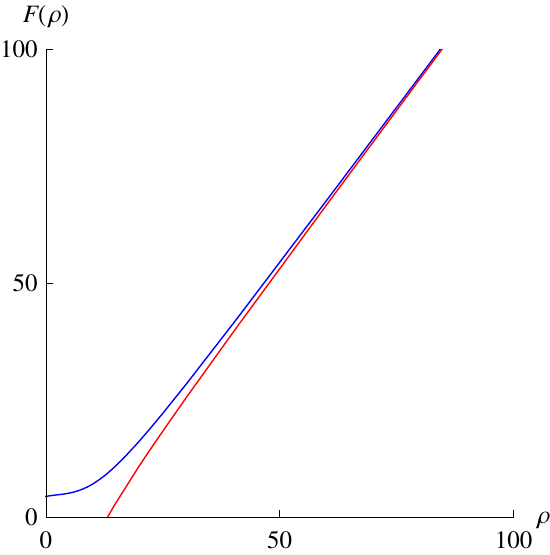}}                
  \subfloat{\label{fig:gNcNf41G2UV}\includegraphics[width=0.333\textwidth]{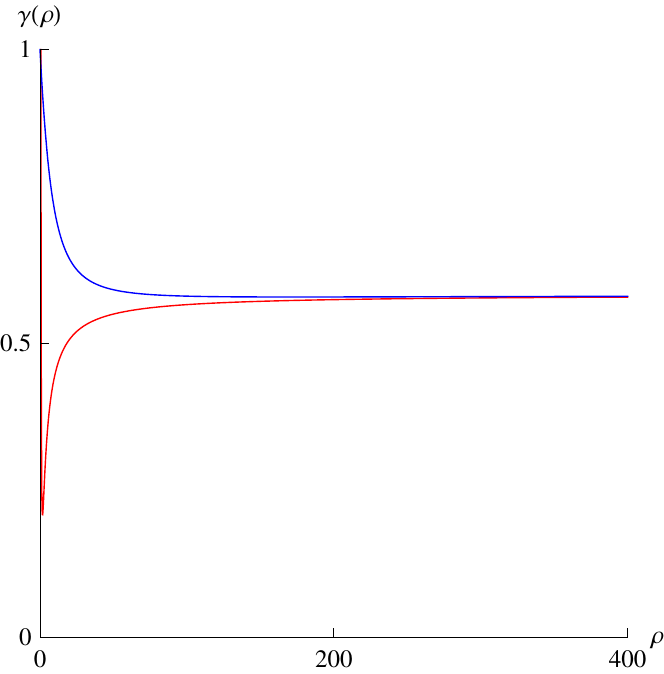}}
  \subfloat{\label{fig:wNcNf41G2UV}\includegraphics[width=0.333\textwidth]{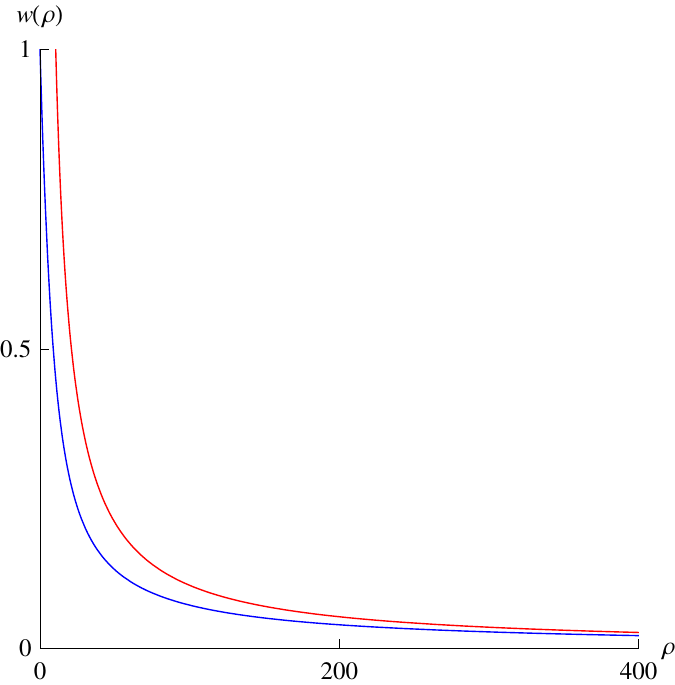}}
  \caption{Numerical matching of asymptotically constant dilaton background for $N_c=4$ and $N_f=1$. The red lines
  are graphs of the solution from the UV expansion while the blue line is the numerical solution. The IR matching is much the same as in \autoref{fig:lindilmatching}.}
  \label{fig:G2matching}
 \end{center}
\end{figure}
We conclude that we have a background that in the IR is the deformed Maldacena-Nastase background of \autoref{section:MN} and in the UV is the massless flavour background of appendix \ref{massless}. With the choice of profile defined in \autoref{eq:profile} we have shown that it is possible to smoothly interpolate between the two, so our background is well defined. In fact numerical matching does not seem to depend on the choice of $C$ in \autoref{section:Ansatz}, indeed the cases $C=-1$ and $C=0$ have been checked.
\subsection{On the Field Theory}\label{section:QFT}
Now that we have a well defined background, at least numerically, we should start to analyse the specifics of the dual field theory. In this section we shall concentrate on the solutions with an asymptotically linear dilaton. Some of the details are clearly going to be the same as the massless case considered in \cite{Canoura:2008at}. In particular for $N_c \geq 2N_f$ the Yang-Mills coupling $g^2_{\text{YM}}\to\infty$ in the IR ($\rho<<m_q$) of the theory and $g^2_{\text{YM}}\to0$ in the UV ($\rho>>m_q$) which is what we expect form a confining theory with asymptotic freedom\footnote{This result can be easily extracted from the $\mathcal{F}^2$ term in the action of a space time filling probe D5 brane wrapping $\Sigma$ as in \cite{Canoura:2008at}}. For $N_c <2N_f$ the location of the IR and UV of the theory in terms of $\rho$ is opposite (UV is $\rho<<m_q$ and vice-versa), in this case the authors of \cite{Canoura:2008at} inferred that this theory developed a Landau pole in the UV. The case $N_c \geq 2N_f$ is both the easier to interpret and the more phenomenologically interesting so it is this we shall concentrate on here. 

\subsubsection{Wilson Loops}\label{section:Loops}
We would like to show that the dual field theory exhibits confinement in the IR and and confirm asymptotic freedom in the UV, at least for $N_c\geq2N_f$. To this end we shall use our supergravity solution to calculated the expectation value Wilson loops in this section.
For an arbitrary path $\mathcal{C}$ a Wilson loop is defined by the following gauge invariant equation:
\begin{equation}
W_{\mathcal{C}}=\frac{1}{N_c}tr\mathcal{P}e^{i\oint A}
\end{equation}
Which is the path-ordered exponential of the gauge field $A$.
In this section we shall calculate two such Wilson loops, the rectangular loop and the circular one.

In euclidean space the expectation value of a rectangular Wilson Loop is related to the quark-antiquark potential $E$. If a rectangular loop, in the $(x,t)$ plane, is defined by $-\frac{L}{2}\leq x\leq\frac{L}{2}$ and $0\leq t \leq T$ then in the limit $T\to\infty$ the relation is given by: 
\begin{equation}
<W_{\mathcal{C}}>\propto e^{-TE(L)}
\end{equation}

In \cite{Maldacena:1998im} Maldacena argued that in order to calculate a Wilson loop from a gravity dual one should consider a minimal surface attached to the loop that extends from the UV into the IR. If this surface has symmetry such that its parametrisation is effectively 2-dimensional we need only minimise the Nambu-Goto action for an open string with ends fixed in the UV. The quark-antiquark potential is then extracted from the minimal Nambu-Goto action through the identification $E=\frac{S_{\text{N.G}}}{T}$.

\begin{figure}
 \begin{center}
  \subfloat[{\footnotesize$P=1$, $C=-1$, $\kappa=\frac{1}{2}-\frac{3N_f}{2N_c}$}]{\label{fig:sqlmassless}\includegraphics[width=0.333\textwidth]{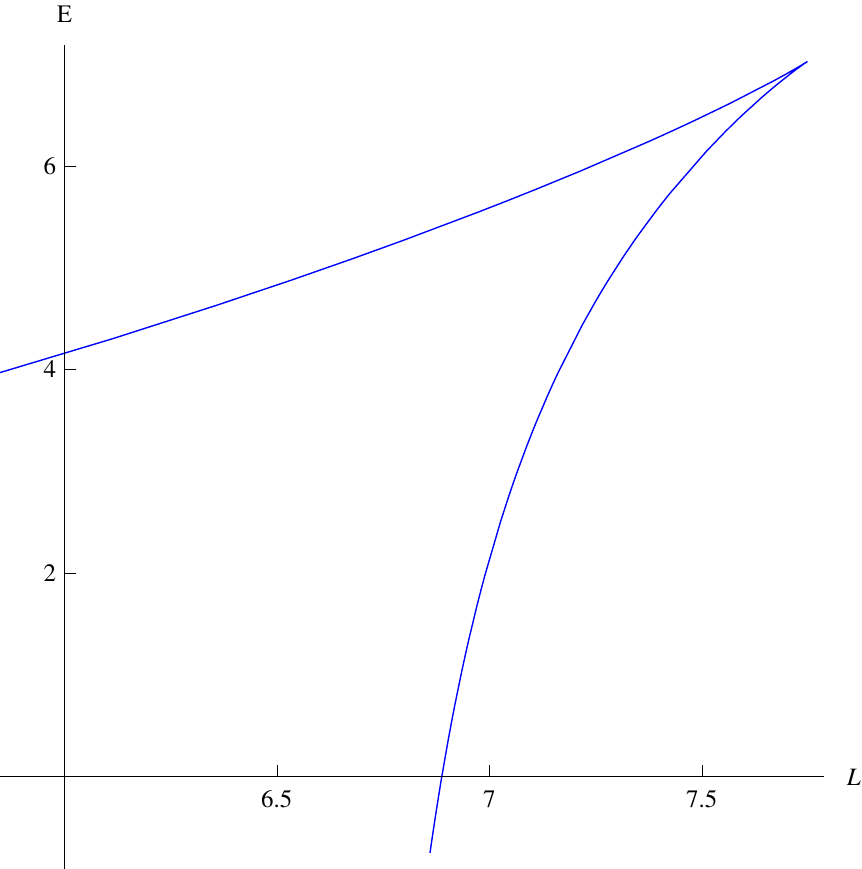}}               
\subfloat[$P=\left(\frac{\pi}{2}\arctan\rho\right)^3$, $C=1$, $\kappa=\frac{1}{2}$]{\label{fig:sqlbetweenless}\includegraphics[width=0.333\textwidth]{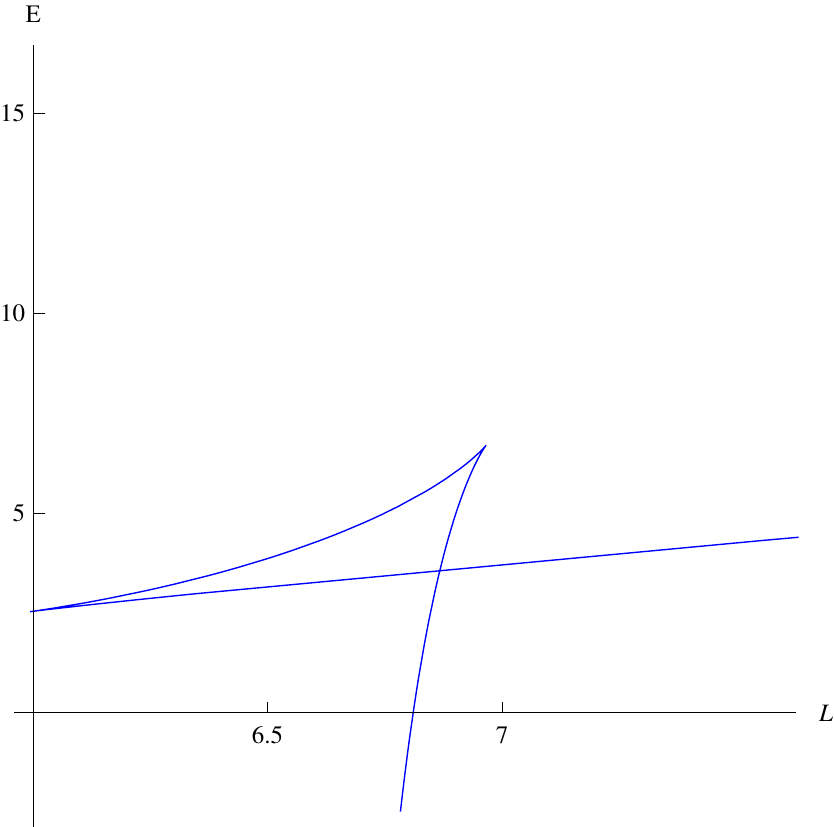}}    \subfloat[$P=\left(\frac{\pi}{2}\arctan\rho\right)^6$, $C=1$, $\kappa=\frac{1}{2}$]{\label{fig:sqlmassive}\includegraphics[width=0.333\textwidth]{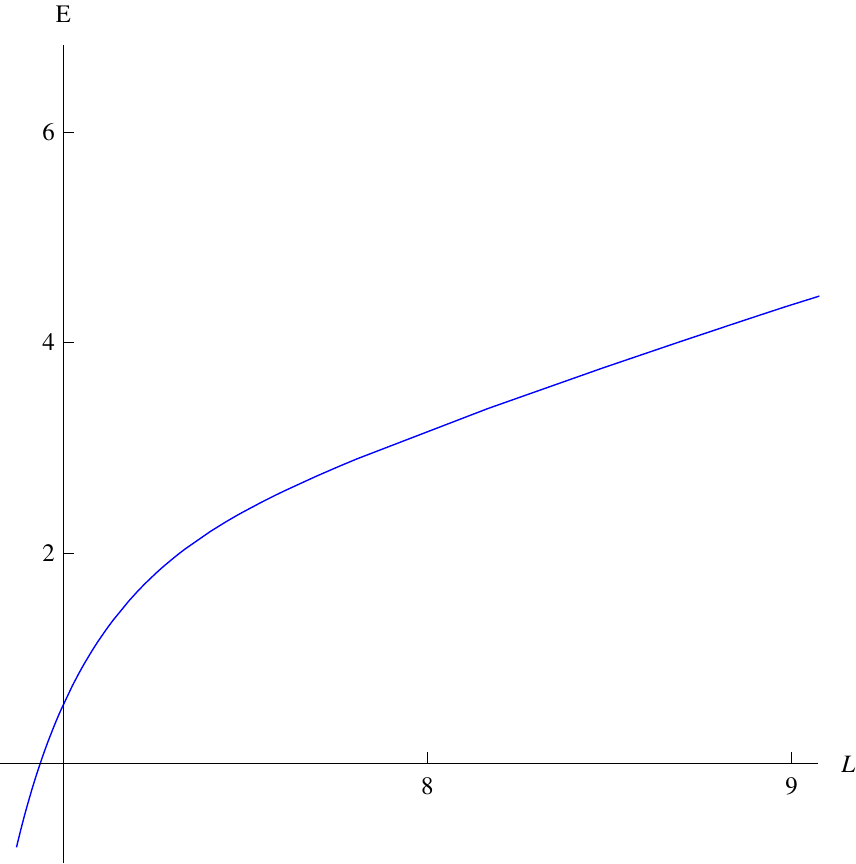}}
  \caption{Here are plots of the energy-length relationship for rectangular Wilson loops with $N_c=4$ and $N_f=1$. (a) shows the massless flavour theory of \cite{Canoura:2008at} where there is a cusp due to the IR singularity. In (b) there is a first order phase transition, like that of the Van der Waals gas. This explicitly shows the the transition between the IR and UV regimes of the theory. (c) Is a plot with the "good" profile of \autoref{section:Matching} which corresponds to a larger common quark mass, $m_q$ than (b). In this case the graph is smooth with no discontinuities and is thus entirely physical.} 
  \label{fig:sql}
 \end{center}
\end{figure}

If we choose to parametrise the world sheet of the string by $\rho(x)$ then for $\chi(\rho)^2=(\frac{dr}{d\rho})^2$ and $\rho'=\frac{d\rho}{dx}$ the induced metric in the sting frame is given by:
\begin{equation}
e^{\phi}\bigg[-dt^2+(1+\chi^2\rho'^2)dx^2\bigg]
\end{equation}
And so the Nambu-Goto action is given by\footnote{see \cite{Nunez:2009da} for a rigorous derivation of what follows for general metrics}:
\begin{equation}
S_{\text{Rectangle}}= T\int_{-\frac{L}{2}}^{\frac{L}{2}} dx e^{\phi}\sqrt{1+\chi^2\rho'^2}
\end{equation}
Because there is no explicit dependence on $x$ in integrand we don't need to derive the second order equations of "motion" but can use the first integral formula to derive the following first order O.D.E:
\begin{equation}
\chi^2\rho'^2=e^{2(\phi-\phi_{\text{min}})}-1
\end{equation} 
Where $\phi_{\text{min}}=\phi(\rho_{\text{min}})$ and $\rho_{\text{min}}$ is the distance of the lowest point of the string from $\rho=0$. From here it is a simple matter to derive an expression for L as a function of $\rho_{\text{min}}$:
\begin{equation}\label{eq:squareLoopL}
L(\rho_{\text{min}})=2\int_{\rho_{\text{min}}}^{\rho_{\text{max}}} d\rho\chi(\rho) \frac{e^{\phi_{\text{min}}}}{\sqrt{e^{2\phi(\rho)}-e^{2\phi_{\text{min}}}}}
\end{equation}
And for the energy of the string as a function of $\rho_{\text{min}}$:
\begin{equation}\label{eq:squareLoopE}
E(\rho_{\text{min}})=e^{\phi_{\text{min}}}L(\rho_{\text{min}}) +2\int_{\rho_{\text{min}}}^{\rho_{\text{max}}}d\rho\chi(\rho) \sqrt{e^{2\phi(\rho)}-e^{2\phi_{\text{min}}}} -2\int_0^{\rho_{\text{max}}}d\rho\chi(\rho)e^{\phi(\rho)}
\end{equation}
Where the last term in \autoref{eq:squareLoopE} is present because we need to subtract the contribution to the Energy from strings that stretch straight from the UV to the IR. In both \autoref{eq:squareLoopL} and \autoref{eq:squareLoopE} formally we should take $\rho_{\text{max}}\to\infty$, however since we cannot do this numerically it is necessary to take a finite upper bound to plot graphs. In \autoref{fig:sql} there are graphs of $E$ as a function of $L$ derived numerically, we can see that for the massive flavour theory we no-longer have the cusp which appears in the massless case. In Fig.\autoref{fig:sqlmassive} we have linear behaviour for large L and it appears that what was interpreted as string breaking due to pair production of quarks in \cite{Canoura:2008at} is a symptom of the IR singularity of the massless background. There are two distinct scales in this theory, in the UV the potential exhibits an inverse power law and in the IR where quarks have been integrated out and the theory has the confining behaviour of deformed Maldacena-Nastase. Fig.\autoref{fig:sqlbetweenless} explicitly shows the first order phase transition between these regimes. This behaviour, as described in \cite{Nunez:2009da}, is what we expect in a theory with two scales, and was first observed in the context of Wilson loops in theories with massive flavour in \cite{Bigazzi:2008gd}.  

To confirm the numerical results we should perform some analytic calculations to determine the precise relation between $E$, $L$ and $\rho_{\text{min}}$, \cite{Nunez:2009da} lays out a procedure to do this\footnote{It should be reiterated that our background is non singular. The often encountered problems with calculating Wilson loops in theories with flavour, discussed in \cite{Nunez:2009da}, will not apply here. Thus we should be able to apply the flavourless techniques laid out in that paper to our current case.}. 
Our goal is to exploit the exact differential relation derived in \cite{Nunez:2009da} and which takes the following form for our background:
\begin{equation}\label{eq:diffe}
\frac{dE}{dL}= e^{\phi(\rho_{\text{min}})}
\end{equation}
This will only be helpful for the asymptotic values of $\rho$ where, as shown in \autoref{section:Solutions}, we have analytic solutions to the background fields in terms of series. The lowest contribution to \autoref{eq:squareLoopL} is of the form:
\begin{equation}\label{eq:Lmin}
L\sim\int\limits_{\rho_{\text{min}}\approx 0} \frac{1}{\rho} =\log(\rho_{\text{min}})\bigg\lvert_{\rho_{\text{min}}=0}
\end{equation}
Which is clearly infinite. The upper limit of the integral is given by:
\begin{equation}
L\sim\int^{\infty}d\rho\sqrt{\rho}e^{-\frac{\rho}{2N_c-4N_f}}
\end{equation}
This shows that that $L$ becomes infinite as $\rho_{\text{min}}\to 0$ provided $N_c>2N_f$, so that the upper bound is finite and cannot cancel the infinite contribution from the lower bound. The expression \autoref{eq:Lmin} can now be inverted to give us $\rho_{\text{min}}(L)$ which then enables the integration of \autoref{eq:diffe}, giving the result:
\begin{equation}\label{eq:En}
E=e^{\phi_{0}}L +O\left(e^{-\sqrt{\frac{7}{24F_0^3}}N_c L}\right)
\end{equation}
This linear behaviour supports the assertion that the dual QFT is confining.

\begin{figure}
 \begin{center}
  \subfloat{\label{fig:sqlmasslessd}\includegraphics[width=0.333\textwidth]{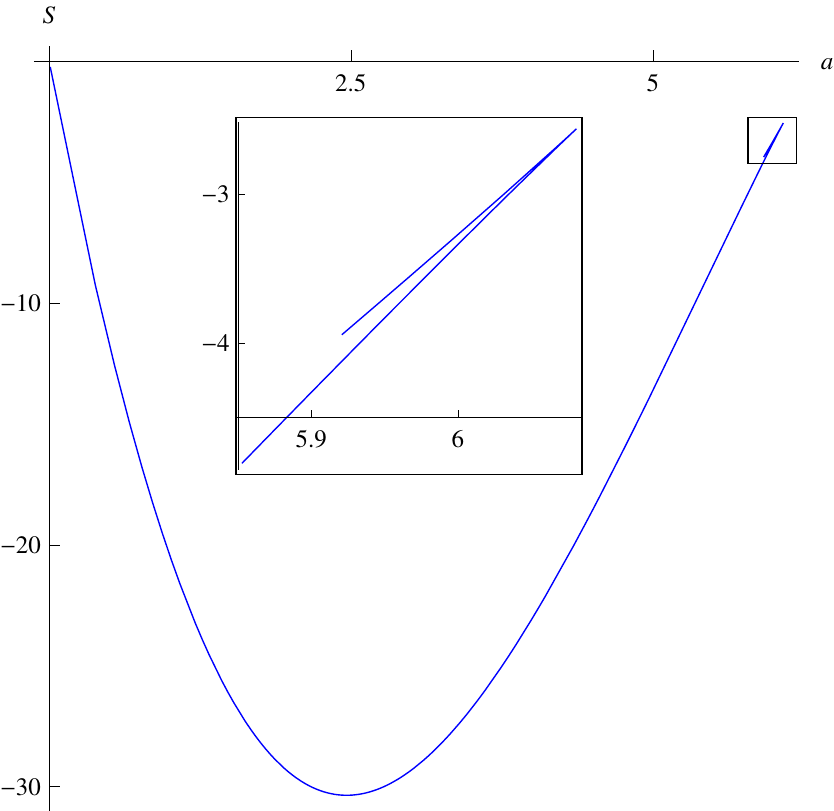}}~~~~~~~~~~~                
  \subfloat{\label{fig:sqlmassived}\includegraphics[width=0.333\textwidth]{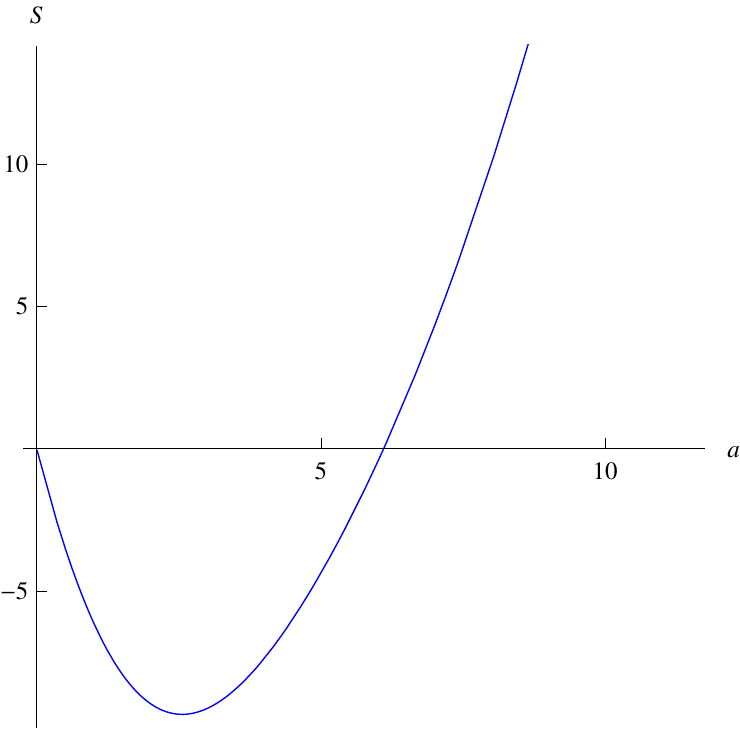}}
  \caption{Graphs of the Nambu-goto actions for circular Wilson loops with $N_c=4$ and $N_f=1$ as a function of a. The left is for the massless case the right for the massive. Notice that the massless loop again has a cusp in the IR.}
   \label{fig:Cir}
 \end{center}
\end{figure}

We now turn our attention to the circular Wilson loop, its expectation value is once more extracted from the Nambu-Goto action of an appropriately parametrised string. In Euclidean space the induced string frame metric of a string constrained to end on a circle in the field theory coordinates for $\rho\to\infty$ is given by:
\begin{equation}
e^{\phi}\bigg[(R^2 d\psi^2 +dR^2) +\chi^2d\rho^2\bigg]
\end{equation}
where the bracketed part is the metric of an arbitrary circle in $(t,x,y)$. If we choose to parametrise the string world sheet by $R(\rho)$ and make the reasonable assumption that it has no angular dependence we are led to the following action:
\begin{equation}\label{eq:CircleLoop}
S_{\text{Circle}}=2\pi\int_{\rho_{min}}^{\rho_{max}} d\rho R(\rho) e^{\phi(\rho)}\sqrt{\chi(\rho)^2+R'(\rho)^2}
\end{equation}
This action explicitly depends on $\rho$ so there is no first integral equation and because our background is not AdS we are unable to call upon the power of conformal transformations to make the sort of simplification performed in \cite{Berenstein:1998ij}. Instead we must calculate by brute force. \autoref{eq:CircleLoop} is minimises by the following differential equation:
\begin{equation}\label{eq:diffR}
R''(\rho)=\frac{\left(R'(\rho )^2+\chi (\rho )^2\right) \left(\chi (\rho )^2-R(\rho ) R'(\rho ) \phi '(\rho )\right)+R(\rho ) \chi (\rho ) R'(\rho ) \chi '(\rho )}{R(\rho ) \chi (\rho )^2}
\end{equation}
This equation is complicated so we must resort to solving it numerically. With $R(\rho)$ determined the expectation value of the Loop can be ascertained. Once more we need to renormalise \autoref{eq:CircleLoop} to subtract the contribution from strings that stretch straight along $\rho$. So if we call the radius of the circular Wilson loop $a$ its expectation value is expressed in terms of the following corrected action:
\begin{equation}\label{eq:corrCir}
\displaystyle \frac{1}{2\pi}S= \int_{\rho_{min}}^{\rho_{max}} d\rho\left( R(\rho)e^{\phi(\rho)}\sqrt{\chi(\rho)^2+R'(\rho)^2}-a e^{\phi(\rho)}\chi(\rho)\right)-a\int_{0}^{\rho_{min}}d\rho  e^{\phi(\rho)}\chi(\rho)
\end{equation}
In \autoref{fig:Cir} there is a numerical plot of the action of the circular loop as a function of its radius, a. For large a the graph becomes quadratic, which it should to be consistent with the result for rectangular loop \autoref{eq:En}. To verify this we preform another semi analytic calculation. First note that if we chose to parametrise \autoref{eq:CircleLoop} in terms of $\rho(R)$, then integrand of the action is given by:
\begin{equation}
\frac{dS_{\text{Circle}}}{dR} = 2\pi R e^{\phi(R)}\sqrt{1+\chi(R)^2\rho'(R)^2}\approx 2\pi R e^{\phi_{0}}
\end{equation}
The approximation on the RHS is good when we consider loops that probe the deep IR. This is because the minimal surface attached to such loops will be close to cylindrical. Upon integrating the approximation we arrive at an area law:
\begin{equation}
S_{\text{Circle}} \approx e^{\phi_{0}}\pi a^2 
\end{equation}
This is consistent with result for the rectangular loop. Indeed we have now shown that in both cases the expectation value of the large loop only depends on the area the loop encloses:
\begin{equation}
<W_{\mathcal{C}}>\propto e^{-e^{\phi_{0}}\text{Area}(\mathcal{C})}
\end{equation}
Thus we conclude that the dual gauge theory exhibits confinement in the IR.

We have shown that we have a good supergravity dual of a QFT with a confining IR, in fact through the semi analytic calculations, we have shown that this is true for any profile consistent with the criteria of \autoref{section:Solutions} which is a robust result. It should be noted that the qualitative results of this section seem not to depend on the choice of $C$ in \autoref{section:Ansatz}, indeed a detailed study has confirms this for $C=0$ and $C=-1$

\section{Massive Flavoured Type IIA Solution}\label{section:IIA}
In \cite{Gaillard:2010gy} Gaillard and Martelli present a solution generating method which gives a new type-IIA supergravity solution in terms of a simpler one. It is an algorithm often referred to as a "Rotation" as this is how it acts on the space of killing spinors. In this section we will apply this rotation to the solutions thus far presented in \autoref{section:Solutions} and \autoref{section:Matching}). As these are for type-IIB, we will have to make use of S-duality along the way. First however we should familiarise ourselves with how the rotation works.

\subsection{Rotation}
In \cite{Gaillard:2010gy} the 11-dimensional supergravity set up of \cite{Martelli:2003ki} is dimensionally reduced to give a metric and supersymmetry conditions for a type-IIA supergravity with an interpolating $G_2$ structure\footnote{Interpolating in the sense of the $SU(3)$ structure of \cite{Gaillard:2010qg}, not in the sense of a flavour profile} preserving  $\mathcal{N}=1$ SUSY on fractional NS5 branes. It has a warped metric given, in the string frame, by:
\begin{equation}
ds^2_{str}=e^{2\Delta+2\phi/3}\left(dx_{1,2}^2+ds_7^2\right)
\end{equation}
Where $ds_7^2$ is a manifold of $G_2$ holonomy. This background is augmented by two non trivial fluxes, $F_{(4)}$ and $H_{(3)}$, in addition to a function $\zeta$   that acts as a phase. The supersymmetry conditions for this system are expressed in terms of the following differential relations involving the, yet to be determined, G-structure form, $\Phi$, which is a 3-form:
\begin{equation}\label{eq:BigBPS}
\left.\begin{array}{l l}
&\Phi\wedge d\Phi=0\\
&d(e^{6\Delta}*_7\Phi)=0\\
&d(e^{2\Delta+2\phi/3}\cos\zeta)=0\\&2d\zeta-e^{-3\Delta}\cos\zeta d(e^{3\Delta}\sin\zeta)=0\\
&\frac{1}{\cos^2\zeta}e^{-4\Delta+2\phi/3}*_7d(e^{6\Delta}\cos\zeta\Phi)=H_{(3)}\\
&Vol_3\wedge d(e^{3\Delta}\sin\zeta)-\frac{\sin\zeta}{\cos^2\zeta}e^{-3\Delta}d(e^{6\Delta}\cos\zeta\Phi)=F_{(4)}\\
\end{array}\right.
\end{equation} 
This is clearly going to lead to a set of highly none trivial differential BPS equations which will be difficult to solve directly. This is why it is advantageous to employ a solution generating method such as rotation. 

It is easiest to elucidate the connection between \autoref{eq:BigBPS} and Maldacena-Nastase with massive flavours by considering the limit $\zeta=0$. The SUSY conditions truncate dramatically, now $F_{(4)}=0$ and the 4th equation implies $2\Delta = -2\phi/3$, at least up to a constant which can be set to zero through a redefinition of $\phi$. We are left with:
\begin{equation}\label{eq:NSBPS}
\left.\begin{array}{l l}
&\Phi\wedge d\Phi=0\\
&d(e^{-2\phi}*_7 \Phi)=0\\
&e^{2\phi}*_7d(e^{-2\phi}\Phi)=H_{(3)}\\
&ds^2_{\text{str}}=dx_{1,2}^2 +ds_7^2\\
\end{array}\right.
\end{equation}
The only non zero flux is now $H_{(3)}$ and if we choose $ds^2_7$ to be:
\begin{equation}\label{equation:ds7}
ds_7^2=  dr^2 +\frac{e^{2h}}{4}(\sigma^i)^2 +\frac{e^{2g}}{4} (\omega^i-A^i)^2 
\end{equation}
We are in fact left with the S-dual of our original D5 brane set up. To be precise, \autoref{eq:NSBPS} is the string frame, S-dualised versions of \autoref{eq:Sforms} and \autoref{eq:Metric}.

The key point of the rotation procedure is that it is possible to find a solution of \autoref{eq:BigBPS}, for any $\zeta$ in terms of \autoref{eq:NSBPS}, where $\zeta$ is zero\footnote{Indeed the name should now be clear as we are effectively rotating from $\zeta=0$ to some other arbitrary value.}. If $\Phi^{(0)}$,  $\phi^{(0)}$ and $H_{(3)}^{(0)}$ solve \autoref{eq:NSBPS} then we can define:
\begin{equation}
\left.\begin{array}{l l}
\vspace{1 mm}
~\hat{\Phi}&=\left(\frac{\cos\zeta}{c_1}\right)^3\Phi^{(0)};~~~~~~ e^{3\hat{\Delta}}=\left(\frac{c_1}{\cos\zeta}\right)^2e^{-\phi^{(0)}};\\
e^{2\hat{\phi}}&=~\frac{\cos\zeta}{c_1}e^{2\phi^{(0)}};~~~~~~~~~d\hat{s}_7^2 =\frac{\cos^2\zeta}{c_1^2}ds_7^{(0)2} ;\\
&~~~~~~~~~~~~~\sin\zeta=c_2e^{-\phi^{(0)}};\\
\end{array}\right.
\end{equation} 
Where the quantities denoted by a hat will satisfy the first 3 equations in \autoref{eq:BigBPS}. The 4th equations is solved when the final relation holds and $c_1$, $c_2$ are integration constants. The fluxes can then also be written in terms of the unrotated quantities and the whole system can be written as:
\begin{equation}\label{eq:rot1}
\left.\begin{array}{l l}
\vspace{1 mm}
d\hat{s}_{str}^2&=H^{-1/2}dx_{1,2}^2+H^{1/2}ds_7^{(0)2};~~~~ H=\frac{1}{c_1^2}\left(1-c_2^2 e^{-2\phi^{(0)}}\right)\\
\vspace{1 mm}
\hat{H}_{(3)}&=\frac{1}{c_1}e^{2\phi^{(0)}}*_7^{(0)}d(e^{-2\phi^{(0)}}\Phi^{(0)})~~~~~~e^{2\hat{\phi}}=e^{2\phi^{(0)}}H^{1/2}\\
&~~~~~\hat{F}_{(4)} = \frac{1}{c_2}Vol_3\wedge dH^{-1}-\frac{c_2}{c_1}d(e^{-2\phi^{(0)}}\Phi^{(0)})\\
\end{array}\right.
\end{equation}
Any solution to \autoref{eq:BigBPS}, with the appropriate Bianchi identities for the fluxes, will be a solution to the type-IIA supergravity equations of motion. The Bianchi identities for the rotated system follow immediately form the unrotated solutions \cite{Gaillard:2010gy}, thus finding a solution of \autoref{eq:NSBPS} implies a new rotated solution of the form \autoref{eq:rot1}.

\subsection{A New Type-IIA Solution}
With the rotation procedure laid out we would now like to apply it to the Maldacena-Nastase with Massive flavours to generate a new $\mathcal{N}=1$ type-IIA solution. Thus far we have dealt only with Type IIB with a non trivial RR 3-form in the Einstein frame. The rotation works for type-IIA in the string frame with NS 3-form, but this is easily reconciled. We change \autoref{eq:Metric} to the string frame via $ds^2_{\text{string}}=e^{\phi/2}ds^2$ and then if we perform an S-duality, it will map the theory into the common Type II NS-sector. The maps are as follows:
\begin{equation}\label{eq:maps}
\phi\to-\phi~~~~F_{(3)}\to H_{(3)}~~~~ds^2_{\text{string}}\to e^{-\phi}ds^2_{\text{string}}=ds^2_{\text{str}}
\end{equation}
Thus \autoref{eq:Metric} will lose its warp factor and we are left with a string frame metric precisely of the form in \autoref{eq:NSBPS}. With the appropriate substitutions in \autoref{eq:rot1} it is  possible to express the new type-IIA solution in terms of the unrotated type-IIB solution as:
\begin{equation}\label{eq:rot}
\begin{split}
&d\hat{s}_{str}^2=H^{-1/2}dx_{1,2}^2+H^{1/2}ds_7^2,~~~~ H=\frac{1}{c_1^2}\left(1-c_2^2 e^{2\phi}\right)\\
&H_{(3)}=-\frac{1}{c_1}F_{(3)},~~~~~~~~~~~~~~~~~~~~ e^{2\hat{\phi}}=e^{-2\phi}H^{1/2}\\
&F_{(4)}=\frac{1}{c_2}Vol_3\wedge dH^{-1}+\frac{c_2}{c_1}e^{2\phi}*_7 F_{(3)}
\end{split}
\end{equation}
Here the Hodge dual is now taken with respect to \autoref{equation:ds7} and $c_1$, $c_2$ are integration constants. We can express these as:
\begin{equation}\label{equation:c1c2}
c_1=-\frac{1}{\cosh\beta},~~~~ c_2=-e^{-\phi_{\infty}}\tanh\beta
\end{equation}
This enables us to relate the rotation to the following chain of transformations: A lift to M-theory, rescaling the 11th dimension by $e^{-\phi_{\infty}}$, a boost along $x_{11}$ with parameter $\beta$, undo the rescaling of $x_{11}$, reduce back to IIA and finally performing two T-dualities along the spatial field theory directions as described in \cite{Gaillard:2010gy}.

The reality of the new metric requires that $H>0$ which imposes $e^{2(\phi-\phi_{\infty})}\tanh^2\beta <1$, thus the unrotated dilaton must be bounded above for the rotated solution to make sense\footnote{This also ensures that the relation $\sin\zeta =c_2e^{\phi}=-\tanh\beta e^{\phi-\phi_{\infty}}$ can hold for any position in space}. It is then the asymptotically constant dilaton (\autoref{eq:constdil}) that is compatible with this rotation. A glance at \autoref{eq:rot} shows that the form of the dilaton has changed. This is because the rotation introduces D2-branes which the dilaton couples to\footnote{See \cite{Caceres:2011zn} for an in depth discussion on related issues}
. This tells us that we must have D2-branes in the rotated system but we may have more and we can only tell by looking at the explicit terms of $F_{(4)}$.

In order to succinctly express the explicit structure of the forms expressed in  \autoref{eq:rot} it is helpful to introduce the following vielbein basis for the rotated theory:
\begin{equation}
\left.\begin{array}{l l}
e_R^{x^i}=H^{-1/4}dx^i,~~~~~~& e_R^r = H^{1/4} dr,\\
e_R^i= H^{1/4} e^{h}\frac{\sigma^i}{2},~~~~~& e_R^{\hat{i}} =H^{1/4}e^{g}\left(\frac{\omega^i - A^i}{2}\right)
\end{array}\right.
\end{equation}
The components of the NS 3-form expressed in this basis are:
\begin{equation}
\left.
  \begin{array}{l l}
  \vspace{3 mm}
H^{(3)}_{\hat{i}jk} =  \frac{N_c}{c_1H^{3/4}}\epsilon_{ijk}\left(1+w^2-\frac{4N_f}{N_c}P-2w\gamma\right)e^{-3h};&
H^{(3)}_{123} = - \frac{N_c}{4c_1H^{3/4}}V e^{-3h};\\
  \vspace{3 mm}
H^{(3)}_{ri\hat{i}} =  -\frac{N_c}{2c_1 H^{3/4}}\left(\frac{4N_f}{N_c}\eta P' +\gamma'\right)e^{-g-h};&H^{(3)}_{\hat{1}\hat{2}\hat{3}} = \frac{2N_c}{c_1 H^{3/4}}e^{-3g};\\
  \vspace{3 mm}
H^{(3)}_{\hat{i}\hat{j}k} =  \frac{N_c}{c_1H^{3/4}}\epsilon_{ijk}\left(w-\gamma\right)e^{-g-2h};&\\
\end{array}\right.
\end{equation}
Where:
\begin{equation}
V=\left(1-w^2\right)\left(w-3\gamma\right) - 4\left(1-\frac{3N_f}{N_c}P\right)w +8\left(\frac{1}{2}+\frac{3N_f}{2N_c}P\right)
\end{equation}
And $\eta$ is defined in \autoref{eq:eta}. The components of the RR 4-form are the following:
\begin{equation}\label{equation:ComponentsofF4}
\left.
  \begin{array}{l l }
    \vspace{3 mm}
F^{(4)}_{r123} = -\frac{2c_2N_c}{c_1H}e^{-3g+2\phi};&F^{(4)}_{r\hat{i}\hat{j}k} = - \frac{c_2N_c}{2c_1H}\epsilon_{ijk}\left(1+w^2-\frac{4N_f}{N_c}P -2w\gamma\right)e^{g-h+2\phi};\\
  \vspace{3 mm}
F^{(4)}_{r\hat{1}\hat{2}\hat{3}} = -\frac{c_2N_c}{4c_1 H} V e^{-3h+2\phi};&F^{(4)}_{r\hat{i}jk} =\frac{c_2N_c}{c_1H}\epsilon_{ijk}\left(w-\gamma\right)e^{-2g-h+2\phi};\\
  \vspace{3 mm}
F^{(4)}_{txyr} = \frac{2c_2}{c1^2 H^{3/2}}\phi'e^{2\phi};
&F^{(4)}_{i\hat{i}j\hat{j}} = -\frac{c_2N_c}{2 c_1H}\left(\frac{4N_f}{N_c}\eta P'+\gamma'\right)e^{-g-h+2\phi};\\
\end{array}\right.
\end{equation}
It is clear $F_{(4)}$ consists of 2 distinct parts. The first has support over the field theory coordinates which implies an electric coupling to D2-branes and the second with no time dependence which implies a magnetic coupling to D4-branes. Thus we see that the rotation has generated D2, D4 and NS5-branes. We can actually go further, from the definition of $F_{(4)}$ in \autoref{eq:rot1} we can see that we can easily extract $C_3$. We just need to make the observation that $\Phi = e^{3f}\Phi^{(0)}$ where $\Phi$ is the G-structure of Appendix \ref{BPS} and $\Phi^{(0)}$ is that of \autoref{eq:NSBPS}. Thus if we take into account the maps in \autoref{eq:maps}, we see that we must be able to express the 3-form gauge field $C_{(3)}$ as:
\begin{equation}
C_{(3)} = -\frac{1}{c_2 H}Vol_{(3)}+\frac{c_2}{c_1}e^{2\phi-3f}\Phi
\end{equation}
It is a simple matter to verify (with Mathematica) that our BPS equations imply $dC_{(3)}=F_{(4)}$ and so the fluxes obey the following equations of motion:
\begin{equation}
dH_{(3)}= -\frac{1}{c_1}4\pi^2\Omega_s;~~~~~~~dF_{(4)}=0
\end{equation}
Thus the rotation has not created any additional sources and the the proof in \cite{arXiv:0706.1244} ensures that all the equations of motion are satisfied.

This background is the $G_2$ analogue of the Baryonic branch, \cite{Butti:2004pk}, of Klebanov-Strassler \cite{Klebanov:2000hb}\footnote{Which is generated by a type-IIB rotation of the Maldacena-Nunez solution \cite{hep-th/0008001}} with additional flavour as in \cite{Gaillard:2010qg}. This suggests that the that the field theory dual is likely to be a quiver gauge theory with additional fundamental flavour group. In \cite{Gaillard:2010gy}, when $P=0$, they note that $C_{(3)}$ is running on a 3 sphere at infinity which is analogous to the $B_2$ field in \cite{Klebanov:2000hb}. The same happens here, indeed on the shrinking 3-cycle  $\Sigma=\{\sigma^i|\sigma^i=\omega^i\}$ the integral of $C_{(3)}$ takes the following asymptotic form with respect to the solutions of \autoref{section:Solutions}:
\begin{equation}
c(\rho)=-\frac{1}{4\pi^2}\int_{\Sigma}C_{(3)}= \left\{
\begin{array}{l l}
\vspace{1 mm}
\frac{N_c}{16F_0^{3/2}}\rho^2-\frac{N_f(F_0-N_c)P_2}{40F_0^{3/2}}\rho^{5/2}+...&\quad \rho\approx0\\
\frac{N_c}{2}\rho^{1/2}-\frac{9(18N_c^3-25N_cN_f-33N_f^2)-28N_fP_1}{96\rho^{1/2}}+...&\quad\rho\approx\infty\\
\end{array}\right.
\end{equation}
We will now set about calculating the charges of the branes in the background to try and gain some information about the field theory dual. When $P\neq0$ care needs to be taken here as we have a source for the NS 3 form $H_{(3)}$, this implies that the usual improved field strengths may not be gauge invariant.

\subsection{The Charge of the Branes}
Clearly, when the $P\neq0$, there is a source for the NS 3-form which is difficult to interpret. We will not attempt to do so here but will calculate the brane charges to get some idea of what is going on in the background. 

The NS5 charge is obtained by integrating $H_{(3)}$ over the 3-sphere parametrised by $\omega^i$. This will of course give a quantisation condition like that obtained for the unrotated D5-brane system in \autoref{equation:D5Quantisation}. However we have an additional factor now, expressed in terms of \autoref{equation:c1c2} this gives:
\begin{equation}
-\frac{1}{4\pi^2}\int_{\tilde{S}^3}H_{(3)}=N_c\cosh\beta = N_c^{(1)}
\end{equation}
where it is now $N_c^{(1)}$ that is quantised.

To calculate the charge on the D2-brane it useful to first make the following observation. From the definitions of $H_{(3)}$, $C_{(3)}$ and $F_{(4)}$ it is possible to show that the following relationship is implied by the BPS system:
\begin{equation}\label{equation:SimpleRelationship}
*_{10}F_{(4)} + e^{2\phi}H_{(3)}\wedge C_{(3)}= \frac{c_2}{32c_1^2}\phi'(1+e^{2\phi})e^{3g+3h}\sigma^1\wedge \sigma^2\wedge\sigma^3\wedge\omega^1\wedge\omega^2\wedge\omega^3
\end{equation} 
Where the term on the right hand side comes entirely from:
\begin{equation}\label{eq:CommomTerms}
(*_{10}F_{(4)})^{\text{Int}}= (H_{(3)}\wedge C_{(3)})^{\text{Int}}=\frac{c_2}{32c_1^2}\phi'e^{3g+3h+2\phi}\sigma^1\wedge \sigma^2\wedge\sigma^3\wedge\omega^1\wedge\omega^2\wedge\omega^3
\end{equation}
The Page and Maxwell charges coincided for the NS5-brane, this is not so for the D2-brane (see \cite{Marolf:2000cb} for a discussion on the various types of charge in a theory with SuGra Chern-Simons terms). The Maxwell charge is gauge invariant but not quantised in general, it is calculated by integrating $*_{10}F_{(4)}$ over a 6-cycle, from a glance at \autoref{equation:ComponentsofF4} it is clear that the only suitable cycle will be $\mathcal{C}_6=\sigma^1\wedge\sigma^2\wedge\sigma^3\wedge\omega^1\wedge\omega^3\wedge\omega^3$. We are thus lead to:
\begin{equation}
N_{D2}=\frac{1}{(2\pi)^5}\int_{\mathcal{C}_6}*_{10}F_{(4)}= \frac{\pi}{4}\phi'\sinh\beta\cosh\beta e^{3g+3h+2\phi-\phi_{\infty}}
\end{equation}
In terms of the $\rho=e^{2h}$ expansions of \autoref{section:Solutions}, where it should be reiterated that it is the asymptotically constant dilaton UV expansion that should be used, we can express this as:
\begin{equation}
\frac{N_{D2}(\rho)}{\frac{\pi}{4}\sinh\beta\cosh\beta e^{\phi_{\infty}}}=\left\{
\begin{array}{l l}
\vspace{1 mm}
\frac{7N_c^2+36N_fF_0^3P_2}{384F_0^{3/2}}\rho^2+...&\quad \rho\approx 0\\
\frac{N_f}{8}\rho^{3/2}+\frac{4N_c^2-9N_cN_f+N_f(3N_f+4P_1)}{32}\rho^{1/2}+...&\quad \rho\approx \infty\\
\end{array}\right.
\end{equation}
The number of D2-branes grows quickly in the UV, which is indicative of the rotation generating an irrelevant operator in the field theory dual as in \cite{Elander:2011mh}\cite{Conde:2011aa}, we will come back to this issue in the next section. One can define Page for the D2 branes by integrating $*F_{(4)}-H_{(3)}\wedge C_{(3)}$ over the same cycle and as shown in \autoref{eq:CommomTerms} we must clearly have that:
\begin{equation}
N_c^{(2)}=\frac{1}{(2\pi)^5}\int_{\mathcal{C}_6}\left(*_{10}F_{(4)}-H_{(3)}\wedge C_{(3)}\right)=0
\end{equation}
The point here is that while $N_c^{(2)}$ is quantised it is not gauge invariant for large gauge transformations, where large means those that cannot be expressed as $\delta C_{(3)} = d\Lambda_{(2)}$ \footnote{Such a gauge transformation will give $\delta(H_{(3)}\wedge C_{(3)})= H_{(3)}\wedge d\Lambda_2$ and since $H_{(3)}$ has a source this might lead one to question whether the effect of $\delta C_{(3)}$ on the integrand of $N_c^{(2)}$ is to add a total derivative. This is the case though because the only possible, non zero, infinitesimal gauge transformations on $\mathcal{C}_6$ are of the form $\Lambda_2^{ij}= \sigma^i\wedge\omega^j$ and $d \Lambda_2^{ij}\wedge dH_{(3)} = 0$ on $\mathcal{C}_6$.} Usually these large gauge transformation, which induce quantised shifts in Page charge, are associated with Seiberg dualities as in \cite{Benini:2007gx}\cite{Aharony:2009fc}. However here it is $C_3$ rather than $B_2$ that is running, and supergravity is not invariant under unit shift in the integral of $c(\rho)$. This complicates issues if we wish to interpret the running of the $N_{D2}$ as a duality cascade.

The D4 Maxwell charge is given by integrating $F_4$ over a compact 4 cycle. It is only the term $F_{ij\hat{i}\hat{j}}^{(4)}$ in \autoref{equation:ComponentsofF4} that does not contract with $dr$ but this does depend on $r$. This implies that the charge is not quantised, it runs. In terms of the $\rho=e^{2h}$ expansions it takes the form:
\begin{equation}\label{eq:D4MCharge}
N_{D4}(\rho)=-\frac{1}{(2\pi)^3}\int_{\mathcal{M}_4} F_{(4)}\propto\left\{
  \begin{array}{l l}
  \vspace{1 mm}
 \frac{N_c}{48F_0^{1/2}}\rho+...& \quad \rho \approx 0\\
 \frac{N_fP_1}{12\rho^{1/2}}+...& \quad  \rho \approx \infty\\
  \end{array}\right.
\end{equation}
Clearly the number of D4-branes vanishes in both the far IR and far UV but is running in between. There is an issue here, the fact that there is no $C_1$ gauge field in this system tells us that Maxwell and Page charges should coincide for the D4-brane. However we see from \autoref{eq:D4MCharge} that even when $P=0$ this charge is running. 

\subsection{A decoupling limit}
At present it is clear that the warp factor $H$ tend to a non-zero constant in the UV (see \autoref{eq:rot}), we would like to define a decoupling limit where the metric become asymptotically AdS so we need $H\to0$ in the UV. $H = (1-\tanh^2\beta e^{2\phi-2\phi_{\infty}})\cosh^2\beta$ so we need to take the limit $\beta\to\infty$ in a well defined way. We proceed as in \cite{Gaillard:2010gy} by rescaling the field theory coordinates:
\begin{equation}
d^2x_{1,2}\to N_c^{(1)} N_c \cosh\beta d^2x_{1,2}
\end{equation}
We can now define a new metric:
\begin{equation}
ds_{str}^2 =N_c^{(1)}\bigg[\tilde{H}^{-\frac{1}{2}}d^2x_{1,2} + \tilde{H}^{\frac{1}{2}}d^2s_7\bigg]
\end{equation}
Where $d^2s_7$ is unchanged but the new warp factor, in the limit $\beta\to\infty$, is given by $\tilde{H}=\frac{1-e^{2\phi-2\phi_{\infty}}}{N_c^2}$. Although this seems like a reasonably sensible limit for the metric, this is not so for $F_{(4)}$ and the rotated dilaton in \autoref{eq:rot}. A finite limit can be obtained if we simultaneously take:
\begin{equation}
\beta\to\infty; ~~~ e^{-\phi_{\infty}}\to0; ~~~~ e^{-2\phi_{\infty}}N_c\sinh\beta \to 1
\end{equation}
Then the fluxes and dilaton are finite as in \cite{Gaillard:2010gy}, however there is a problem with $\tilde{H}$. From the asymptotically constant dilaton expansions (\autoref{eq:G2series1},\autoref{eq:constdil}), we have that:
\begin{equation}
\tilde{H}\sim\frac{9N_f}{2N_c^2\rho} +\frac{9(4N_c^2+12N_cN_f+N_f(4P_1-9N_f))}{16N_c^2\rho^2}
\end{equation}
The leading order term here represents a deformation away from the field theoretically acceptable AdS asymptotics. This is due to the sharply rising number of D2-branes in the UV which must be inducing D2-brane charge on the NS5-branes in an analogous way to what happens in \cite{Conde:2011aa}. There it was a growing number of D3-branes on the conifold, where it signalled a departure of the dual field theory from 4-dimensional cascade behaviour dual to an irrelevant operator. We expect the same picture to apply here, in particular, as in \cite{Conde:2011aa}, a profile which decays for large $\rho$ will remove the irrelevant operator and give the field theory dual a UV completion. We can achieve this by introducing a profile of the form $P\sim\frac{P_1}{\rho}$ for large $\rho$, this is equivalent to setting all $N_f$ terms which do not multiply a $P_i$ to zero in the series solutions \autoref{eq:G2series1}, \autoref{eq:constdil}\footnote{Numerical matching  was confirmed for the test profile $P=\frac{1}{\rho}(\frac{2}{\pi}\arctan\rho)^6$} .For such a profile we have that in the UV:
\begin{equation}
N_{D2}\sim\frac{(N_c^2+N_fP_1)\sqrt{\rho}}{8};~~~~~\tilde{H}\sim\frac{9(N_c^2+N_fP_1)}{4N_c^2\rho^2}
\end{equation}
This asymptotic behaviour does indeed lead to a decoupling limit. Expressed in terms of $\rho$, when $P\sim\frac{P_1}{\rho}$, the leading order terms of the other functions in the metric are given by:
\begin{equation}
\frac{dr}{d\rho}\sim \sqrt{\frac{3}{4\rho}};~~~~ e^{2g}=F\sim\frac{4\rho}{3}~~~~ e^{2h}=\rho
\end{equation}
Which implies that after a further rescaling of the field theory coordinate, the metric can be expressed as
\begin{equation}\label{eq:ads}
d^2s_{str}=\frac{9N_c^{(1)}\sqrt{1+N_fP_1/N_c^2}}{2}\bigg[U^2~ d^2x_{1,2}+\frac{dU^2}{U^2} +\frac{(\sigma^i)^2}{12} +\frac{(\omega^{i}-\frac{1}{2}\sigma^{i})^2}{9}\bigg]
\end{equation}
Where we have introduced a new holographic coordinate $U=\sqrt{\rho}$. This explicitly shows that, when $\beta\to\infty$, the metric is asymptotically $AdS_4\times Y$ where Y is the metric at the tip of the Bryant-Salamon $G_2$ cone \cite{Bryand:1989mv}. Thus we have defined a decoupling limit, $\rho\to\infty$, where the the asymptotically Minkowski region can be removed an replaced by a boundary. The theory is not however asymptotically conformal as the rotated dilaton has the expansion:
\begin{equation}
e^{2\tilde{\phi}}\sim \frac{3\sqrt{N_c^2+N_fP_1}}{2N_c^2\rho}
\end{equation}

\section{Discussion}\label{section:diss}
In this work we considered adding unquenched massive fundamental matter to the Maldacena-Nastase background \cite{Maldacena:2001pb}. This was done by introducing a flavour profile that interpolates between the deformed Maldacena-Nastase solution and a Massless flavoured solutions similar to the one found in \cite{Canoura:2008at}. It was explicitly shown that, starting from a profile of a quite arbitrary form, there are asymptotic, semi-analytic, solutions to the BPS system. These are consistent massive flavour deformations of both the linear dilaton and flavoured $G_2$-cone solutions found in \cite{Canoura:2008at}. The massive BPS system is free from the IR singularities that so plague backgrounds with massless flavours (see \cite{Conde:2011sw} for an exception) and in particular it was possible to derive a semi analytic solution in the IR. A phenomenologically motivated profile was proposed to interpolate between the asymptotic solutions and it was shown via a numerical study that, indeed, the IR and UV expansions can be continuously connected. The pathology free background was then shown to be dual in the IR to $\mathcal{N}=1$ SYM-CS with level $k=\frac{N_c}{2}$ and $N_f$ massive flavours. There is a degree of ambiguity in defining the Chern-Simons level away from the deep IR. We decided to have level that is quantised everywhere so that the system is consistent after rotation. It would be interesting to see whether there is a field theory explanation for the levels $N_f$ independence, or whether one has to abandon quantisation of k over the whole range of the holographic coordinate. However we found that the qualitative field theory results did not depend to much on this choice. It would also be interesting to see if this theory has a mass gap. A recent field theory calculation \cite{Agarwal:2012bn} has shown that a gap exists for pure $\mathcal{N}=1$ SYM-CS but the effect of additional fundamental matter is, at least to our knowledge, yet to be explored. For the holographic dual, the dilaton and warp factors of the metric are bounded in the IR which is suggestive of gap but it would be interesting to confirm this with a calculation of the glue ball spectrum, in the spirit of \cite{Witten:1998zw}\cite{Caceres:2005yx}.

On the formal side, it is unfortunate that a derivation of a profile corresponding to a specific massive brane embedding is absent here. It was derived for the dual of SQCD with massive flavours in \cite{Conde:2011rg}. They achieved this via a microscopic calculation for a simplified case and then, using a mathematical trick, they conjecture a profile for the general case. Sadly neither technique seems to be applicable for this background, at least for now, and the reasons are two fold. Firstly the microscopic calculation relies heavily on the fact that their internal space is the conifold, here it is a $G_2$-manifold and the mathematical machinery, applicable to the former, is absent. Secondly a partial integration of the BPS system allows a vast simplification of the conifold system, such a simplification for $G_2$-manifold is yet to be derived. We leave the resolution of these issues to future work and trust that profiles do exist for concrete brane embeddings. 

We also employed a solution generating technique called "rotation", \cite{Gaillard:2008wt}, to generate a new type-IIA supergravity solution. This is the $G_2$ analogue of Baryonic branch of Klebanov-Strassler \cite{Butti:2004pk} but with a profile. It seems reasonable to expect that, as with the system before rotation, the dual field theory after rotaion will have a Chern-Simons level $k=\frac{N_c}{2}$, however the non trivial $F_{(4)}$ means we cannot perform the same easy check via s-duality. There are various difficulties with interpreting the field theory dual of this solution. The Page charge on the D4-branes is running (even when $P=0$), there is a source for the NS5-branes, a running $C_3$ and a pile up of D2-branes in the UV . We offer no explanation for the running Page charge other than that Page charge quantisation may not be universal. The source for the NS5-branes is unusual and there are perhaps issues with defining an action that includes this, however at the level of the equations of motion there does not seem to be any a priori problem with such a source. The picture of invariance of the integral of $B_2$ under shifts by $1$, which is associate with a chain of Seiberg dualities in a duality cascade in \cite{Benini:2007gx} does not follow clearly here. However since we do have running integral of $C_3$ and the D2-brane Page charge is quantised it is reasonable to expect similar behaviour, indeed in some loose sense this background is the T-dual of certain type-IIB conifold backgrounds that exhibit such behaviour. The fast increasing number of D2-branes in the UV is by now well understood in terms of the rotation creating an irrelevant operator in the UV of the field theory \cite{Caceres:2011zn}\cite{Elander:2011mh}\cite{Conde:2011aa}. It signals the departure of the dual field theory from $2+1$ dimensional dynamics and indicates the need for a UV completion. This issue was solved, as in \cite{Conde:2011aa}, by using a profile that decays in the UV and we where in fact able to show that it is possible to take a decoupling limit in which the background becomes asymptotically $AdS_4 \times Y$, where $Y$ is the tip of the Bryant-Salamon $G_2$ cone \cite{Bryand:1989mv}. It seems reasonable that the dual field theory will be a $2+1$ dimensional quiver with gauge group $SU(N_c^{(2)})\times SU(N_c^{(1)}+N_c^{(2)} +\frac{n_f}{2})$, where $n_f=\cosh\beta N_f$ and $N_c^{(2)}=c N_c^{(1)}$, for $c$ the integral of $C_3$. This fits nicely with the results \cite{Gaillard:2010qg}, where sources are added to branes on the conifold. Unfortunately such a convincing set of tests, as exists for \cite{Gaillard:2010qg}, are not yet available for this background so we leave the confirmation of the speculative gauge group for future work.

\section{Acknowledgements} 
I am indebted to Carlos N\'u\~nez for the many useful discussions during this project, his knowledge and insight where a great help to me. I would also like to thank my supervisor Adi Armoni for guidance and encouragement and I am grateful to them both for help improving this manuscript. I would also like to thank Alfonso Ramallo and J\'er\^ome Gaillard for useful comments. My work is supported by an STFC studentship.

\appendix
\section{Derivation of the BPS equations}\label{BPS}
In what follows the following vielbein basis is used:
\begin{equation}\label{eq:vielbeins}
\left.\begin{array}{l l} 
e^{x^i}=e^fdx^i,~~~~~~& e^r = e^f dr,\\
e^i= e^{f+h}\frac{\sigma^i}{2},~~~~~& e^{\hat{i}} =e^{f+g}\left(\frac{\omega^i - A^i}{2}\right)
\end{array}\right.
\end{equation}
The projections which preserve $\mathcal{N}=1$ SUSY for the massless flavour case are given as\cite{Canoura:2008at}:
\begin{equation}\label{eq:projections}
i\epsilon^*=\epsilon,~~\Gamma_{1\hat{1}}\epsilon=\Gamma_{2\hat{2}}\epsilon=\Gamma_{3\hat{3}}\epsilon,~~ \Gamma_{\hat{1}\hat{2}\hat{3}}\epsilon= (\cos\alpha + \sin\alpha\Gamma_{1\hat{1}})\epsilon
\end{equation}
Where $\epsilon$ is a killing spinor satisfying $\bar{\epsilon}\epsilon = 1$ and $\alpha = \alpha(r)$ is an angle to be determined. We require that the massive flavour background also satisfies these equations. Using these projections it is possible to derive the BPS equations as a consequence of the vanishing SUSY variations of the dilatino and gravitino. An alternative route to the BPS equations is to use G-structures\cite{Gauntlett:2003cy}. The internal space is a manifold of $G_2$ Holonomy and so it is equipped with an associative 3-form defined in terms of the fermionic biliners by:
\begin{equation}
\Phi_{a,b,c}= -i \bar{\epsilon}\gamma_{a,b,c}\epsilon
\end{equation}
Where $\gamma_a$ is a 7 dimensional gamma matrix. For a background to be supersymetric the 3-form, $\Phi$, must satisfy the following set of equations:
\begin{equation}\label{eq:Sforms}
\begin{split}
e^{-3f-\phi}d(e^{3f+\phi/2}\Phi) = &~*_7 F_3\\
d(e^{\phi}*_7\Phi)= &~0\\
\Phi\wedge d\Phi = &~0\\
d(e^{2f-\phi/2}) = &~0
\end{split}
\end{equation}
For the specific projections obeyed by this background the $G_2$-Structure form can be shown to be given by: 
\begin{equation}\label{eq:PhiJ}
\Phi = e^r\wedge J - \text{Re}(e^{-i\alpha}\Omega)
\end{equation}
Where:
\begin{equation}
\begin{split}
J = & \sum_{i=1}^{3} e^i\wedge e^{\hat{i}}\\
\Omega = & (e^1 + ie^{\hat{1}})\wedge(e^2 + ie^{\hat{2}})\wedge(e^3 + i e^{\hat{3}})
\end{split}
\end{equation}
Specifically this gives:.
\begin{equation}\label{eq:G2st}
\begin{array}{l l}
\Phi= & e^r \wedge\left( e^1 \wedge e^{\hat{1}} + e^2 \wedge e^{\hat{2}} + e^3 \wedge e^{\hat{3}}\right)\\ &-\cos\alpha\left( e^1 \wedge e^2 \wedge e^3 - e^{\hat{1}} \wedge e^{\hat{2}} \wedge e^3 - e^{\hat{1}} \wedge e^2 \wedge e^{\hat{3}}-e^1\wedge e^{\hat{2}}\wedge e^{\hat{3}} \right)\\ &+ \sin\alpha\left(e^{\hat{1}} \wedge e^{\hat{2}} \wedge e^{\hat{3}} - e^{\hat{1}} \wedge e^2 \wedge e^3 - e^1\wedge e^{\hat{2}} \wedge e^3- e^1\wedge e^2 \wedge e^{\hat{3}}\right)
\end{array}
\end{equation}
Inserting this 3-form into \autoref{eq:Sforms} and performing some algebra (with Mathematica) gives the following set of first order differential BPS equations:
\begin{equation}\label{eq:BPS}
\left.
  \begin{array}{l l}
\vspace{3 mm}   
  \phi'= & 4 f'\\ 
f'= & \frac{1}{32} e^{-3g-2h} \sec\alpha\big(8e^{2h}N_c-12e^{3g+h}\sin^2\alpha w +\\
\vspace{3 mm}
   &6e^{4g}\sin^2\alpha(w^2-1)-6e^{2g}(4e^{2h}\sin^2\alpha+N_c(1-L_2+w^2-2 w \gamma))\big)\\

g'= &\frac{1}{4} e^{-3 g-2 h} \cos\alpha \big(e^{2 g} \left(-L_2 N_c+w N_c
   (w-2 \gamma )+N_c+4 e^{2 h}\right)-4 e^{2 h} N_c+\\
   \vspace{3 mm} 
   &e^{4 g}\left(w^2-1\right)\big)+N_c e^{-2 g-h} (w-\gamma ) \sin \alpha\\

h'= & \frac{1}{8} e^{-3 g-2 h} \sec \alpha \big(4
   e^{g+h} \sin^2\alpha \left(w \left(2
   e^{2 g}-N_c\right)+N_c \gamma \right)+8 e^{2 h}
   \big(3 e^{2 g}-\\
\vspace{3 mm}   
   &N_c\big) \sin ^2\alpha +e^{2 g}
   (\cos 2 \alpha -5) \left(e^{2 g}
   \left(w^2-1\right)+N_c \left(L_2+2 w \gamma
   -w^2-1\right)\right)\big)\\

w'= & e^{-4 g} \big(N_c e^{2 g-h} \sin \alpha
   \left(-L_2-2 w \gamma +w^2+1\right)+4 e^{h}
   \\
\vspace{3 mm}  
   &\left(2 e^{2 g}-N_c\right) \sin \alpha+4 e^{g}
   \cos \alpha  \left(w \left(e^{2
   g}-N_c\right)+N_c \gamma \right)\big)\\
\vspace{3 mm}
\gamma'= &-L_1+ \frac{e^{-h} \sin \alpha  \left(4 e^{2
   h}-e^{2 g} \left(w^2-1\right)\right)}{N_c}+\frac{4
   e^{g} w \cos \alpha}{N_c}\\

\alpha '= & \frac{3}{4} e^{-3 g-2 h} \sin \alpha  \big(-e^{2 g} \left(-L_2 N_c+w N_c (w-2 \gamma )+N_c+12 e^{2 h}\right)+\\
&4 e^{2 h} N_c+e^{4 g} \left(w^2-1\right)\big)+\frac{3}{4}e^{-3 g-2 h} \cos \alpha  \left(4 N_c e^{g+h} (w-\gamma )-8 w e^{3 g+h}\right)
  \end{array} \right.
\end{equation}
%\begin{equation}\label{eq:BPS}
%\begin{split}
%\phi'= & 4 f'\\
%f'= & \frac{1}{32} e^{-3g-2h} \sec\alpha\left(8e^{2h}N_c-12e^{3g+h}\sin^2\alpha w +\\
%   &6e^{4g}\sin^2\alpha(w^2-1)-6e^{2g}(4e^{2h}\sin^2\alpha+N_c(1-L_2+w^2-2 w \gamma))\right)\\
%
%g'= &\frac{1}{4} e^{-3 g-2 h} \left(e^{2 g} \cos \alpha
%   \left(4 e^{2 h}+N_c (-L_2+N_c w (w-2
%   \gamma)+N_c\right)+\\
%   &4 N_c e^{g+h} (w-\gamma
 %  ) \sin\alpha+e^{4 g} \left(w^2-1\right) \cos \alpha
 %  -4 N_c e^{2 h} \cos\alpha \right)\\
%
%h'= & \frac{1}{8} e^{-3 g-2 h} \sec \alpha \left(4
%   e^{g+h} \sin^2\alpha \left(w \left(2
%   e^{2 g}-N_c\right)+N_c \gamma \right)+8 e^{2 h}
%   \left(3 e^{2 g}-\\
%   &N_c\right) \sin ^2\alpha +e^{2 g}
%   (\cos 2 \alpha -5) \left(e^{2 g}
%   \left(w^2-1\right)+N_c \left(L_2+2 w \gamma
%   -w^2-1\right)\right)\right)\\
%
%w'= & e^{-4 g} \left(N_c e^{2 g-h} \sin \alpha
%   \left(-L_2-2 w \gamma +w^2+1\right)+4 e^{h}
%   \\
%   &\left(2 e^{2 g}-N_c\right) \sin \alpha+4 e^{g}
%   \cos \alpha  \left(w \left(e^{2
%   g}-N_c\right)+N_c \gamma \right)\right)\\
%
%\gamma'= &-L_1+ \frac{e^{-h} \sin \alpha  \left(4 e^{2
%   h}-e^{2 g} \left(w^2-1\right)\right)}{N_c}+\frac{4
%   e^{g} w \cos \alpha}{N_c}\\
%
%\alpha '= & \frac{3}{4} e^{-3 g-2 h} \left(-e^{2 g} \sin
%   \alpha  \left(12 e^{2 h}+N_c (-L_2+N_c
%   w(w-2 \gamma )+N_c\right)+\\
%   &4 N_c e^{g+h}
%   (w-\gamma ) \cos \alpha-8 w e^{3 g+h} \cos
%   \alpha +e^{4 g} \left(w^2-1\right) \sin \alpha +4
%  N_c e^{2 h} \sin \alpha \right)
%\end{Split}
%\end{equation}
In addition to the following algebraic Identity:
\begin{equation}\label{eq:AlgIdnt}
{\textstyle\cot \alpha = \frac{e^{g+h} \left(e^{2 g} \left(1-w^2\right)+4 e^{2
   h}\right)-N_c e^{h-g} \left(e^{2 g} L_2-e^{2 g}
   \left(w^2+1\right)+2 e^{2 g} w \gamma +\frac{4}{3} e^{2
   h}\right)}{+\frac{1}{6} e^{2 g} N_c V-4 w e^{2 g+2 h}+2
   N_c e^{2 h} \left(w-\gamma \right)}\\}
\end{equation}
Where:
\begin{equation}
V=\left(1-w^2\right)\left(w-3\gamma\right) - 4\left(1-\frac{3L_2}{4}\right)w +8\left(\kappa+\frac{3 C L_2}{8}\right)
\end{equation}
From \autoref{eq:AlgIdnt} it is possible to derive the values of the various trigonometric functions contained in \autoref{eq:BPS} and if we take its derivative, after some tedious algebra, we arrive at the following consistency requirement:
\begin{equation}\label{eq:dL2Rule}
L_2'=-\frac{e^{-g} \left(\cos\alpha\left((1-w^2)e^{2g}+4e^{2h}\right)-4w e^{g+h}\sin\alpha \right)}{e^{g}(w+C)\cos\alpha
   +2e^{h}\sin\alpha}\text{L}_1
\end{equation}
Although the proof in \cite{arXiv:0706.1244}, ensures this it is a simple matter to show that this BPS system solve the type-IIB supergravity equations of motion. As a constance check this was done but as it is not particularly illuminating and it shall be omitted in the interest of brevity.

\section{The dual of $\mathcal{N}=1$ SYM in 2+1 dimensions with massless flavours}\label{massless}
In \cite{Canoura:2008at} massless flavours were added to the Maldacena-Nastase background, the purpose of this appendix is give a brief review and to allow comparison of the expansions of massive and massless backgrounds. The starting point is the following Einstein frame metric that was proposed to account for the flavour deformation:
\begin{equation}
ds^2 = e^{2f}\left(dx_{1,2}^2+ dr^2+\frac{e^{2h}}{4} (\sigma^i)^2+\frac{e^{2g}}{4}(\omega^i-A^i)^2\right)
\end{equation}
Where as consequence of the vanishing of the SUSY variations of the dilatino and gravitino we have that $\phi = 4 f$. $A^i$ is chosen as in \autoref{eq:1forms}. The RR 3-form of Maldacena-Nastase gets an extra piece, $f_3$ , so that it now satisfies \autoref{eq:dF3}:
\begin{equation}\label{eq:masslessF3}
F_3 = \hat{F}_3 +f_3
\end{equation}
Where $\hat{F}_3$ is defined by \autoref{eq:MNF3}, $B^i$ chosen as in \autoref{eq:1forms} and the most general flavour modification to the RR 3-form can be written as:
\begin{equation}
f_3 = \frac{N_f}{8}\epsilon_{ijk}\left(\omega^i-(C+1)\frac{\sigma^i}{2}\right)\wedge \sigma^j\wedge\sigma^k
\end{equation}
Where $C$ is a constant\footnote{In \cite{Canoura:2008at} $C=0$ but an arbitrary $C$ changes none of their results.}. This gives the following components for $F_3$ in the vielbein basis (\autoref{eq:vielbeins}):
\begin{equation}\label{eq:coF3massless}
\left.\begin{array}{l l}
F^{(3)}_{\hat{i}jk} = -\frac{N_c}{2}\epsilon_{ijk}\left(1+w^2-\frac{4N_f}{N_c}-2w\gamma\right)e^{-3f-3h};&F^{(3)}_{123} = ~\frac{N_c}{4}V e^{-3f-3h};\\
F^{(3)}_{ri\hat{i}}=~\frac{1}{2}N_c\gamma'e^{-3f-g-h};&F^{(3)}_{\hat{1}\hat{2}\hat{3}} = -2N_c e^{-3f-3g};\\
F^{(3)}_{\hat{i}\hat{j}k} = -N_c\epsilon_{ijk}\left(w-\gamma \right)e^{-3f-g-2h}\\
\end{array}\right.
\end{equation}
Where:
\begin{equation}\label{eq:V0}
V =  \left(1-w^2\right) \left(w-3\gamma\right) -4 \left(1-\frac{3 N_f}{N_c}\right) w
  +8 (\kappa+\frac{3C N_f}{2N_c})
\end{equation}
The pull-back of $F_{(3)}$ onto the cycle on which the color branes are wrapped, $\Sigma$ defined by \autoref{eq:cy}, which must vanish in the IR for a non singular background, then determines the value of the integration constant to be:
\begin{equation}\label{eq:masslesskappa}
\kappa=\frac{1}{2}-\frac{3(C+1)N_f}{2N_c}
\end{equation} 
With this choice of $\kappa$ all dependence of the set up on $C$ then drops out. Observe also that the choice $C=-1$ is equivalent to having $\kappa=\frac{1}{2}$ which is useful to make contact with the massive flavour case where $\kappa$ is forced to take this value.

Irrespective of the choice of $\kappa$, upon taking the exterior derivative of \autoref{eq:masslessF3} we arrive at the following smearing form:
\begin{equation}\label{eq:masslessOmega}
\Omega_s = - \frac{N_f}{8\pi^2}\frac{1}{8} \epsilon_{ilm}\epsilon_{ijk}\sigma^l\wedge\sigma^m\wedge\omega^j\wedge\omega^k
\end{equation}
This smearing form fills the codimensions of a tipple branched supersymmetric brane embedding.

In \cite{Canoura:2008at} the BPS equations were solved in terms of:
\begin{equation}
\rho = e^{2h}~~~~~F = e^{2g}
\end{equation}
Where, for $N_c\geq 2 N_f$, $\rho=0$ corresponds to the IR boundary. A curvature singularity means that a semi analytic, small $\rho$, solution cannot be derive. However on constancy grounds it was argued that the IR values of the various functions in the set up took the following forms:
\begin{equation}\label{eq:fff}
\textstyle F_{\text{IR}}\approx F_0;~~~\gamma_{\text{IR}} \approx 1 - \frac{2N_f}{N_c};~~~w_{\text{IR}} \approx 1
\end{equation}
There is no such problem in the UV where there are in fact 2 expansions, one with an asymptotically linear dilaton \autoref{eq:lineardilaton} and one where the dilaton asymptotes to a constant \autoref{eq:constantdilaton}. In \cite{Canoura:2008at} they decided to proceed numerically from $\rho\approx0$ and showed that the, near to, IR conditions \autoref{eq:fff} could be smoothly connected to the UV expansions, with $\kappa$ defined as in \autoref{eq:masslesskappa}. At this stage it should be pointed out that since the background turns out to be singular anyway it is questionable whether this choice of $\kappa$ is actually required. Indeed \autoref{eq:masslesskappa} is defined in the IR where the effect of the curvature singularity is so severe that a small $\rho$ expansion does not appear to exist. Since $\kappa$ defines the precise form of the UV expansions the choice \autoref{eq:masslesskappa} implies that the unphysical IR singularity is actually dictating what the UV expansion is and this seems perverse. At the very least we should except that other choices of $\kappa$ might well be valid, in the UV at least. All of these other choices leave the $C$ dependence in the set up and it is for this reason that the UV expansions below have been written for arbitrary $\kappa$. 

The UV expansion with asymptotically linear dilaton is given by:
\begin{equation}\label{eq:lineardilaton}
\left.
  \begin{array}{l l}
  \vspace{3 mm}
   			  F=  & {\scriptstyle N_c} + \frac{N_c N_f}{\rho} -\frac{3 N_c N_f \left(N_c - N_f\right)}{4 \rho^2} + \frac{N_c N_f \left(21N_c^2-148N_c N_f+240N_f^2\right)}{16 \rho^3}+ ...\\
   
\vspace{3 mm}

     		     \gamma=  &\frac{2N_c\kappa+3 N_f C}{2\rho} +\frac{5}{8}\frac{(N_c-2N_f)(2N_c\kappa+3N_f C)}{\rho^2}+ \frac{(2N_c\kappa+3N_f C)(49N_c^2-208 N_f N_c +254 N_f^2)}{32\rho^3} +...\\ 
     			
\vspace{3 mm}
     			
    	w=	&  \frac{2N_c\kappa+3 N_f C}{2\rho} +\frac{5}{8}\frac{(N_c-2N_f)(2N_c\kappa+3N_f C)}{\rho^2}+ \frac{(2N_c\kappa+3N_fC)(49N_c^2-184N_f N_c +204N_f^2)}{32\rho^3} +...\\

  \vspace{3 mm}

       \phi=  &\frac{\rho}{2(N_c-2N_f)} +\frac{\left(3 N_c^2-12 N_c N_f+16 N_f^2\right) }{8 (N_c-2 N_f)^2}\log (\rho
   )
   + ...\\
  \end{array} \right.
\end{equation}
While the asymptotically constant dilaton gives rise to the so called flavoured $G_2$-cone solution:
\begin{equation}\label{eq:constantdilaton}
\left.
  \begin{array}{l l}
  \vspace{3 mm}
   			  F = & \frac{4}{3}\rho + {\scriptstyle4(N_f-N_c)} +\frac{N_c^2(15-16\kappa^2)-3N_cN_f(13+16\kappa C)+12N_f^2(2-3C^2)}{\rho}+...\\

\vspace{3 mm}

     		     \gamma= & \frac{2\kappa}{3}+\frac{C N_f}{N_c} + ...~;~ w =  3\frac{(2N_c\kappa + 3N_f C)}{2\rho} + ...\\
     			
\vspace{3 mm}
     			
\phi = &\phi_\infty -\frac{9 N_f}{4\rho}-9\frac{N_c^2(3+4\kappa^2)+\frac{9}{32}N_f^2C^2+6N_c(N_f+2N_f\kappa C)}{\rho^2}   + ...\\
  \end{array} \right.
\end{equation}
%\begin{equation}\label{eq:constantdilaton}
%\begin{split}
%F(\rho) = & \frac{4}{3}\rho + 4(N_f-N_c) +\frac{11N_c^2-15N_cN_f-12N_f^2}{\rho}+...\\
%\gamma(\rho)= & \frac{1}{3}-\frac{N_f}{N_c} + ...\\
%w(\rho) = & \frac{3(N_c - 3N_f)}{2\rho} + ...\\
%\phi(\rho) = &\phi_0 -\frac{9 N_f}{4\rho}-9\frac{4N_c^2+9N_f^2}{32\rho^2}+...
%\end{split}
%\end{equation}

%\bibliography{Mphys Project}
%%
%% Bibliography
%%

\end{document}